\documentclass[conference]{IEEEtran}
\usepackage[utf8]{inputenc}
\usepackage[T1]{fontenc}
\usepackage[unicode]{hyperref}
\usepackage{ifthen}
\usepackage{cite}
\usepackage[cmex10]{amsmath}
\usepackage{amssymb,mathtools,braket,yhmath}
\usepackage{tikz}\usetikzlibrary{arrows.meta,matrix,tikzmark}
	\usetikzlibrary{external}\tikzset{external/export=false}
\usepackage{tikzpagenodes}
\usepackage{lettrine}
\usepackage[at]{easylist}
\usepackage{soul}

\newcommand\fermat[2][]{\ignorespaces
	\tikz[remember picture,overlay]\coordinate(X)at(0,-.5ex);
	\marginpar[{\normalsize\hfill
		\tikz[rp,blue]{
			\node(Y)[fermat=right,#1]{#2};
			\pgfresetboundingbox;
			\draw([yshift=-10]Y.north east)--([yshift=10]Y.south east);}%	
		\tikz[rp,o,blue]{
			\path(Y.east)+(.35,0)|-coordinate(Z)(X);
			\draw(X)+(0,1ex)circle(.05)|-(Z)--(Y);}%
	}]{\normalsize%
		\tikz[rp,blue]{
			\node(Y)[fermat=left,#1]{#2};
			\pgfresetboundingbox;
			\draw([yshift=-10]Y.north west)--([yshift=10]Y.south west);}%
		\tikz[rp,o,blue]{
			\path(Y.west)+(-.35,0)|-coordinate(Z)(X);
			\draw(X)+(0,1ex)circle(.05)|-(Z)--(Y);}%
	}
}
\renewcommand\fermat[2][]{\ignorespaces}
\tikzset{
	every picture/.style={cap=round,join=round},
	rp/.style={remember picture},
	o/.style={overlay},
	fermat/.style={text width=2in,left,align=flush #1},
	b/.style={baseline=#1},
	gb/.style={xscale=5,yscale=3},
	GB/.style={xscale=8,yscale=5},
	bly/.style={below left,yshift=4},
	blx/.style={below left,xshift=4},
	bry/.style={below right,yshift=4},
	arx/.style={above right,xshift=-15},
	->!/.style={dotted,->,shorten >=2},
	F/.style={shape=F},
	E/.style={shape=E,thick,draw,fill=yellow},
	D/.style={shape=D,thick,draw,fill=lime},
	BB/.style={baseline=-.5ex,nodes={inner sep=1}},
	dot/.pic={\fill circle(1pt);},
	tree/.style={nodes={inner sep=1}},
	hl/.style={line width=30/1.44^0#1,cyan},
	hm/.style={line width=30/1.44^0#1,magenta},
}
\makeatletter
\pgfmathsetseed{30717958}
\DeclareUnicodeCharacter{2022}{pic{dot}} % U+2022 = •

\def\readpair#1,#2,{\xdef\xcoord{#1}\xdef\ycoord{#2}}
\tikzdeclarecoordinatesystem{tree}{
	\readpair#1,
	\pgfmathtruncatemacro\yfactor{2^\xcoord}
	\pgfpoint{\xcoord cm}{-5cm*((\ycoord-.5)/\yfactor-.5)}
}
\def\setuptreecs{\let\E\expandafter\def\tikz@parse@regular##1(##2){
	\tikz@parse@cs@tree(##2)\E##1\E{\tikz@return@coordinate}}}

\def\bbsize{8pt}  % bb = building block = black box != blackboard bold
\def\bbwire{12pt} % ≥ \bbsize
\def\bbgape{4pt}  % ≤ \bbsize
\def\bbbase{24pt} % ≥ 2*\bbwire
\def\readtriple#1#2,#3,{\def\xsign{#1}\def\xcoord{#2}\def\ycoord{#3}}
\tikzdeclarecoordinatesystem{bb}{\readtriple#1,
	\pgfpoint{(\xsign(\bbbase*\xcoord+2.24*2^(\xcoord)))}{\ycoord*28}}
\tikzdeclarecoordinatesystem{BB}{\readtriple#1,
	\pgfpoint{(\xsign(\bbbase*\xcoord+1.6*2^(\xcoord)-1))}{\ycoord*20}}
\def\setupbbcs{\let\E\expandafter\def\tikz@parse@regular##1(##2){
	\tikz@parse@cs@bb(##2)\E##1\E{\tikz@return@coordinate}}}
\pgfdeclareshape{F}{
  \savedanchor\NW {\pgfqpoint{-\bbsize}{ \bbsize}}
  \savedanchor\NE {\pgfqpoint{ \bbsize}{ \bbsize}}
  \savedanchor\SW {\pgfqpoint{-\bbsize}{-\bbsize}}
  \savedanchor\SE {\pgfqpoint{ \bbsize}{-\bbsize}}
  \savedanchor\NWI{\pgfqpoint{-\bbsize}{ \bbgape}}
  \savedanchor\NWO{\pgfqpoint{-\bbwire}{ \bbgape}}
  \savedanchor\NEI{\pgfqpoint{ \bbsize}{ \bbgape}}
  \savedanchor\NEO{\pgfqpoint{ \bbwire}{ \bbgape}}
  \savedanchor\SWI{\pgfqpoint{-\bbsize}{-\bbgape}}
  \savedanchor\SWO{\pgfqpoint{-\bbwire}{-\bbgape}}
  \savedanchor\SEI{\pgfqpoint{ \bbsize}{-\bbgape}}
  \savedanchor\SEO{\pgfqpoint{ \bbwire}{-\bbgape}}
  \anchor{center}{\pgfpointorigin}
  \anchor{NW}{\NWO}\anchor{NE}{\NEO}\anchor{SW}{\SWO}\anchor{SE}{\SEO}%
  \backgroundpath{
    \pgfpathrectanglecorners{\SW}{\NE}
    \pgfpathmoveto{\NWI}\pgfpathlineto{\NWO}
    \pgfpathmoveto{\NEI}\pgfpathlineto{\NEO}
    \pgfpathmoveto{\SWI}\pgfpathlineto{\SWO}
    \pgfpathmoveto{\SEI}\pgfpathlineto{\SEO}
  }
}
\pgfdeclareshape{E}{
  \inheritsavedanchors[from=F]
  \anchor{center}{\pgfpointorigin}
  \anchor{NW}{\NWO}\anchor{NE}{\NEO}\anchor{SW}{\SWO}\anchor{SE}{\SEO}
  \inheritbackgroundpath[from=F]
  \foregroundpath{
    \pgftransformscale{.5}\pgfpathmoveto{\SW}\pgfpathlineto{\NE}
  }
}
\pgfdeclareshape{D}{
  \inheritsavedanchors[from=F]
  \anchor{center}{\pgfpointorigin}
  \anchor{NW}{\NWO}\anchor{NE}{\NEO}\anchor{SW}{\SWO}\anchor{SE}{\SEO}
  \inheritbackgroundpath[from=F]
  \foregroundpath{
    \pgftransformscale{.5}\pgfpathmoveto{\NE}\pgfpathlineto{\SW}
    \pgfpathquadraticcurveto{\pgfpointorigin}{\NW}\pgfpathlineto{\SE}
  }
}

\def\Arikan{Ar\i kan }

\def\Bha{Bhattacharyya }
\def\W#1#2{W_{#1}^{(#2)}}
\def\A#1#2{\mathcal A_{#1}^{(#2)}}
\def\AA#1#2{\A{#1}{\<#2\>}}
\def\clA{\mathcal A}
\def\bbB{\mathbb B}
\def\bbE{\mathbb E}
\def\bbF{\mathbb F}
\def\bbN{\mathbb N}
\def\bbP{\mathbb P}

\def\erfc{\operatorname{erfc}}
\def\weight{\operatorname{weight}}
\def\ub{^{\text{upper}}_{\text{bound}}\strut}
\def\lb{^{\text{lower}}_{\text{bound}}\strut}
\def\({\left(}
\def\){\right)}
\def\[{\pgfutil@ifnextchar*{\begin{equation*}\pgfutil@gobble}{\begin{equation}}}
\def\]{\pgfutil@ifnextchar*{\end{equation*}\pgfutil@gobble}{\end{equation}}}
\def\<{\lceil}
\def\>{\rfloor}
\def\sm#1{{\scriptstyle\left[\begin{smallmatrix}#1\end{smallmatrix}\right]}}
\newtheorem{theorem}{Theorem}
\newtheorem{lemma}[theorem]{Lemma}
\newtheorem{corollary}[theorem]{Corollary}

\begin{document}

\title{Polar Code Moderate Deviation: \\
	Recovering the Scaling Exponent}
\author{%
  \IEEEauthorblockN{Hsin-Po Wang and Iwan Duursma}
  \IEEEauthorblockA{University of Illinois at Urbana--Champaign \\
                    \{hpwang2, duursma\}@illinois.edu}
}
\maketitle

\begin{abstract}
	In 2008 \Arikan proposed \emph{polar coding} \cite{Arikan09}
	which we summarize as follows:
	(a) From the root channel $W$ synthesize recursively
		a series of channels $\W N1,\dotsc,\W NN$.
	(b) Select sophisticatedly a subset $\clA$ of synthetic channels.
	(c) Transmit information using synthetic channels indexed by $\clA$
		and freeze the remaining synthetic channels.
	
	\Arikan gives each synthetic channel a ``score'' (called the \Bha parameter)
	that determines whether it should be selected or frozen.
	As $N$ grows, a majority of the scores are either very high or very low,
	i.e.,~they \emph{polarize}.
	By characterizing how fast they polarize, \Arikan showed that
	polar coding is able to produce a series of codes
	that achieve capacity on symmetric binary-input memoryless channels.
	
	In measuring how the scores polarize the relation among
	block length, gap to capacity, and block error probability are studied.
	In particular, the \emph{error exponent regime}
	fixes the gap to capacity and varies the other two.
	The \emph{scaling exponent regime}
	fixes the block error probability and varies the other two.
	The \emph{moderate deviation regime} varies all three factors at once.
	
	The latest result \cite[Theorem~7]{MHU16} in the moderate deviation regime
	does not imply the scaling exponent regime as a special case.
	We give a result that does. (See Corollary~\ref{cor:recover}.)
\end{abstract}

\section{Introduction}

\subsection{The Path to Capacity}

	Assume we want to communicate over the binary erasure channel $W$
	with erasure probability $Z(W)$.
	The Shannon capacity of this channel is $I(W)=1-Z(W)$.
%	\fermat{Instead of ``\Bha parameter'',
%		I simply call it erasure probability.}
	
	Given a series of block codes $\bbB_1,\bbB_2,\dotsc$
	we may calculate their block lengths $N_1,N_2,\dotsc$,
	code rates $R_1,R_2,\dotsc$,
	and block error probabilities $P_1,P_2,\dotsc$.
	An ideal situation is that as $N_n$ goes to infinity, 
	the code rate $R_n$ approaches the channel capacity $I(W)$
	while the block error probability $P_n$ tends to zero.
	This is called \emph{capacity achieving} in the literature.
%	\fermat{Cite Shannon 1948 here?
%		What is the modern definition of capacity achieving??}
	See Fig.~\ref{fig:gb1} for visualization.
	
	Let \emph{gap to capacity} $I(W)-R_n$ be the difference between
	the channel capacity $I(W)$ and the code rate $R_n$.
	There are three factors that we want to understand:
	block length $N_n$, gap to capacity $I(W)-R_n$,
	and the block error probability $P_n$.
	And there are three regimes that study the relation among these factors:
	\emph{error exponent} regime, \emph{scaling exponent} regime,
	and \emph{moderate derivation} regime.
	(See also \cite[Abstract]{MHU16} for a concise summary.)

	\def\axes{
		\path(-.3,-.1);
		\draw[->](0,0)--(0,1)node[o,bly]{Gap to capacity};
		\draw[->](0,0)--(1,0)node[o,blx]{Block error probability};
	}
	\begin{figure}
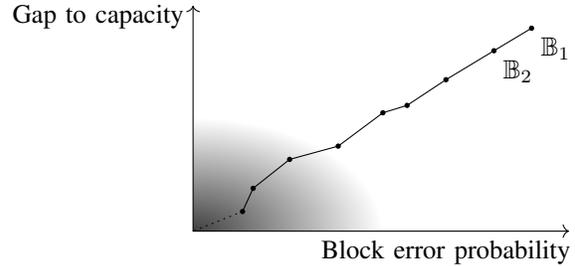
\centering
		\tikz[gb]{
			\begin{scope}
				\clip(0,0)rectangle(1,1);
				\shade[inner color=darkgray](0,0)circle(.5);
			\end{scope}
			\axes
			\draw(.9,.9)•node[o,below right]{$\bbB_1$}
				--(.8,.8)•node[o,below right]{$\bbB_2$}
				foreach\n in{7,...,1}{--(.\n-rand/20,.\n-rand/20)•}
				coordinate(B9);
			\draw[dotted](B9)--(0,0);
		}\caption{The gap to capacity $I(W)-R_n$ ranges from $0$ to $I(W)$;
			the smaller, the better.
			The block error probability $P_n$ ranges from $0$ to $1$;
			the smaller, the better.}
		\label{fig:gb1}
	\end{figure}
	\begin{figure}\centering
		\tikz[gb]{
			\shade[left color=gray,right color=white]
				(0,0)rectangle(.6,.2);
			\draw[o](.02,.2)--node[left]{$I(W)-R\lb$}+(-.04,0);
			\axes
			\draw(.9,.1)•node[o,above right]{$\bbB_1$}
				--(.8,.1)•node[o,above]{$\bbB_2$}
				foreach\n in{7,...,1}{--(.\n-rand/20,.1-rand/20)•}
				coordinate(B9);
			\draw[dotted](B9)--(0,.1);
		}\caption{Error exponent regime:
			Bound the gap to capacity $I(W)-R_n$ from above and measure
			how fast the block error probability $P_n$ goes to zero.}
		\label{fig:gb2}\par\vspace{\intextsep}
		\tikz[gb]{
			\shade[bottom color=gray,top color=white]
				(0,0)rectangle(.2,.6);
			\draw[o](.2,.02)--node[o,below]{$P\ub$}+(0,-.04);
			\axes
			\draw(.1,.9)•node[o,above right]{$\bbB_1$}
				--(.1,.8)•node[o,right]{$\bbB_2$}
				foreach\n in{7,...,1}{--(.1-rand/20,.\n-rand/20)•}
				coordinate(B9);
			\draw[dotted](B9)--(.1,0);
		}\caption{Scaling exponent regime:
			Bound the block error probability $P_n$ from above and
			measure how fast the gap to capacity $I(W)-R_n$ tends to zero.}
		\label{fig:gb3}\par\vspace{\intextsep}
		\tikz[gb]{
			\begin{scope}
				\clip(0,0)rectangle(1,1);
				\shade[inner color=darkgray](0,0)circle(.5);
			\end{scope}
			\axes
			\draw(.9,.9)•node[o,below right]{$\bbB_1$}
				--(.8,.8)•node[o,below right]{$\bbB_2$}
				foreach\n in{7,...,1}{--(.\n-rand/20,.\n-rand/20)•}
				coordinate(B9);
			\draw[dotted](B9)--(0,0);
		}\caption{Moderate deviations regime:
			Measure how fast the series of points $\(P_n,I(W)-R_n\)$
			tends to the origin.}
		\label{fig:gb4}
	\end{figure}
	
\subsubsection{Error Exponent Regime} \label{sec:eer}

	Fix the gap to capacity
	(more precisely, bound the code rate $R_n$ from below)
	and measure how fast the block error probability $P_n$ goes to zero.
	See Fig.~\ref{fig:gb2} for visualization.
	
	For classical polar codes,
%	\fermat{Which one is better:
%		``classical polar codes'' or ``\Arikans polar codes''?}
	$\bbB_n$ is generated by some subset of rows of the tensor power
	$\sm{1&0\\1&1}^{\otimes n}$ and $N_n=2^n$.
	It could be made such that $P_n$ is of order
	\[O\(2^{-N_n^{\beta'}}\)=O\(2^{-2^{n\beta'}}\)\]
	as $n\to\infty$ for all $\beta'<1/2$ and all $R\lb<I(W)$
%	\fermat{We did not define ``$R\lb$'' and its relation with $R_n$.
%		Perhaps it is obvious?}
	\cite[Theorem~1]{AT09}.
	We hence say that the classical polar codes have error exponent%
	\footnote{Remark:
		Gallager \cite{Gallager65}\nocite{Gallager68} proved that
		random codes achieve $P_n=e^{O(N_n)}$.
		Thus in information theory the error exponent is defined differently as
		$\liminf_{n\to\infty}\(-\log P_n\)/N_n$.
		But to distinguish $\beta$ from $\mu$ the scaling exponent,
		we insist on calling $\beta$ the error exponent.
		See also \cite[Abstract]{KSU10}.}
%	\fermat{Clearly the ``exponent'' make sense to
%		a series of codes and a method to generate a series of codes.
%		I will call the former equivalent error exponent in Preliminary.}
	\[\beta\coloneqq\sup_{\text{$\bbB_n$: polar codes}}\liminf_{n\to\infty}
		\frac{\log\(-\log P_n\)}{\log N_n}=\frac12.\]
	
	This characterization is later refined by \cite[Formula~(9)]{HMTU13}
	where $\log_2(-\log_2P_n)$ is
	\[\frac12n+\frac12Q^{-1}\(\frac{R_n}{I(W)}\)\sqrt n+o\(\sqrt n\)\]
	for $Q(\xi)=\erfc\(\xi/\sqrt2\)/2$ the Q-function in statistics.
	
	Their argument applies to generalized polar codes
	that use a larger kernel other than $\sm{1&0\\1&1}$.
%	\fermat{The matrix is tiny?
%		Fix or keep?
%		Fight or flee?}
	The general formula \cite[Formula~(9)]{HMTU13} suggests an obvious obstacle
	$\beta\le1$ since $\beta$ is the average of partial distances divided by
	the block length.
%	\fermat{Do I have to repeat the general formula here?
%		The formula involve partial distance which is hard to define inline.}
	
	In \cite[Example~32]{KSU10} it is given an explicit $16$-by-$16$ kernel
	with error exponent $0.51828$, a number larger than $1/2$.
	They also give a general construction based on Bose--Chaudhuri--Hocquenghem
	codes that achieves error exponents arbitrarily close to $1$,
	as the kernel size grows \cite[Abstract and Section~VI]{KSU10}.
%	\fermat{It is worth notice (or is it obvious?) that one has to sacrifice
%		kernel size to achieve best error exponent.}
	
	For an even more general scenario where the alphabet is $\bbF_q$,
	a similar result is given in \cite{MT14}.
	Specifically Reed--Solomon matrices achieve error exponents arbitrarily
	close to $1$ as the field size (and thus the kernel size) grows.
%	\fermat{It is unclear whether $1$ is achievable for any fixed $q$.
%		This however is minor and I do not want to address here.}
	
	See Appendix~\ref{app:betaruler} for comparison.

\subsubsection{Scaling Exponent Regime} \label{sec:ser}

	Fix (bound from above) the block error probability $P_n$
	and measure how fast the gap to capacity $I(W)-R_n$ tends to zero.%
	\footnote{It is called so because a natural question is
		``what block length $N_n$ do we need
		to achieve a given gap to capacity $I(W)-R\lb$?''
		In reverse the question becomes
		``what gap to capacity $I(W)-R_n$
		can be achieved with block length $N_n$?''
		Perhaps, as the figures illustrate,
		``gap exponent'' is a better name.}
	See Fig.~\ref{fig:gb3} for visualization.
	
	For classical polar codes, \cite[Fig.~2--5]{KMTU10}
	did extensive simulations and suggests that
	$I(W)-R_n$ might, at best, be of order
	\[O\(N_n^{-1/3.6261}\)=O\(2^{-n/3.6261}\)\]
	as $n\to\infty$ for all $P\ub$ and $I(W)$.
%	\fermat{We did not define ``$P\lb$'' and its relation with $P_n$.
%		Perhaps it is obvious?}
	We hence say that classical polar codes ``might'' have
	scaling exponent
	\[\mu\coloneqq\sup_{\text{$\bbB_n$: polar codes}}\limsup_{n\to\infty}
		\frac{-\log(N_n)}{\log\(I(W)-R_n\)}\approx3.6261.\]
	Later \cite{HAU14} provides $\mu\le6$ by a more rigorous reasoning.
	The idea goes as follows: (See also \cite[Formula~(27--33)]{HAU14})
	
	Let $Z_n$ be the \Bha process as in
	\cite[Formula~(67)]{HAU14}.
%	\fermat{\cite{HAU14} uses the term \Bha process''.}
	Let $a>0$ be a number close to $0$ and $b<1$ a number close to $1$.
	The conservation of entropy 
%	\fermat{Reference for conservation of entropy?
%		(i.e.,~the second law of thermodynamics.)}
	suggests that the probability $\bbP(a<Z_n<b)$ controls the gap to capacity.
	Let $g_0$ be the indicator function of the open interval $(a,b)$
	then
	\[\bbE g_0(Z_0)=\bbP(a<Z_0<b).\]
	Define
	\[g_1(\xi)\coloneqq \frac{g_0\(\xi^2\)+g_0\(2\xi-\xi^2\)}2\]
	then
	\begin{align}
		\bbE g_0(Z_1) &= \bbP(a<Z_1<b) \\
			&= \frac{\bbP\(a<Z_0^2<b\)+\bbP\(a<2Z_0-Z_0^2<b\)}2 \\
			&= \frac{g_0\(Z_0^2\)+g_0\(2Z_0-Z_0^2\)}2 \\
			&= g_1(Z_0).
	\end{align}
	Iterate this idea by defining
	\[g_{n+1}(\xi)\coloneqq\frac{g_n\(\xi^2\)+g_n\(2\xi-\xi^2\)}2\]
	to get
	\[\bbE g_0(Z_n)=\bbP(a<Z_n<b)=g_n(Z_0).\]
	The function $2^{n/\mu}g_n$ seems to converge numerically pointwisely
	for some magical choice of $\mu$ \cite[Fig.~5]{HAU14}.
%	\fermat{Like... you literally look at the plot and
%		it ``seems'' to converge.}
	If the limit $g_\infty$ does exist, then
	\[\bbP(a<Z_n<b)=g_n(Z_0)\approx 2^{-n/\mu}g_{\infty}(Z_0).\]
	We summarize the discussion above in the bra-ket notation
	\begin{align}
		\bbP(a<Z_n<b)
			&= \Braket{g_0|Z_n}=\Braket{g_0|T^n|Z_0} \\
			&= \Braket{g_n|Z_0}\approx2^{-n/\mu}\Braket{g_0|Z_0}.
	\end{align}
	
	The consequence is that, as they choose some explicit, machine-handleable
	polynomial to approximate $g_\infty$ they deduce $3.579\le\mu\le6$
	\cite[Abstract]{HAU14}.
	Later in \cite[Fig.~3]{GB14} a more accurate approximation is used
	to obtain the bound $\mu\le5.702$.%
	
	Finally the idea is formulated as following clean criterion.
	\begin{theorem}\cite[Theorem~1 and Formula~(15)]{MHU16} \label{thm:hmu}
		Let $h:[0,1]\to[0,1]$
		be such that $h(0)=h(1)=0$ and $h(\xi)>0$ otherwise.
		If
		\[\sup_{0<\xi<1}\frac{h\(\xi^2\)+h\(2\xi-\xi^2\)}{2h(\xi)}
			<2^{-1/\mu^*} \label{eq:hmu}\]
		for some $\mu^*>2$, then
%		\fermat{This theorem should be stated explicitly
%			due to the nature of Theorem~\ref{thm:gamma}}
		\[\mu\le\mu^*.\]
	\end{theorem}
	\begin{figure}\centering
%		\tikzset{external/export}
		\includegraphics{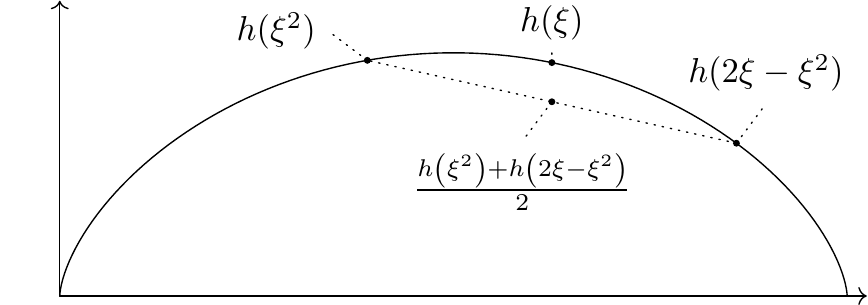}
%		\tikz{
%			\draw plot[raw gnuplot]function{
%				set parametric;x(t)=(cos(t)+1)/2;
%				h(x)=(x*(1-x))**(.64);
%				plot[0:pi]8*x(t),6*h(x(t))smooth csplines;
%			};
%			\draw[dotted]
%				(3.125,2.394)•--+(-.4,.3)node[left]{$h(\xi^2)$}
%				(5,2.371)•--+(0,.1)node[above]{$h(\xi)$}
%				(6.875,1.552)•--+(.3,.4)node[above]{$h(2\xi-\xi^2)$}
%				(6.875,1.552)--(3.125,2.394)
%				(5,1.973)•--+(-.3,-.4)node[below]
%				{$\frac{h\(\xi^2\)+h\(2\xi-\xi^2\)}2$};
%			\pgfresetboundingbox
%			\draw[->](-.6,0)(0,0)--(8.2,0);
%			\draw[->](0,0)--(0,3);
%		}
		\caption{Visualization of $h(\xi)\coloneqq\(\xi(1-\xi)\)^{.64}$
			and how the expected value drops.}
		\label{fig:h}\par\vspace{\intextsep}
		\includegraphics{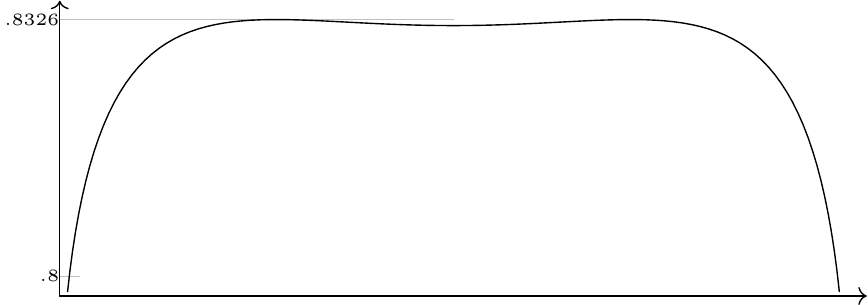}
%		\tikz[nodes={inner sep=0pt}]{
%			\def\a{80}\def\b{63.8}
%			\draw[domain=4:0,samples=2,lightgray]plot function{
%				.8326*\a-\b
%			}node[left,black]{$\scriptscriptstyle.8326$};
%			\draw[domain=.2:0,samples=2,lightgray]plot function{
%				.8*\a-\b
%			}node[left,black]{$\scriptscriptstyle.8$};
%			\draw plot[raw gnuplot]function{
%				set parametric;
%				x(t)=(cos(t)+1)/2;
%				p(x)=x**2;s(x)=2*x-x**2;
%				h(x)=(x*(1-x))**(.64);
%				S(x)=h(p(x))+h(s(x));
%				R(x)=S(x)/2/h(x);
%				plot[.2:pi-.2]8*x(t),R(x(t))*\a-\b smooth csplines;
%			};
%			\pgfresetboundingbox
%			\draw[->](-.6,0)(0,0)--(8.2,0);
%			\draw[->](0,0)--(0,3);
%		}
		\caption{Visualization of $\frac{h\(\xi^2\)+h\(2\xi-\xi^2\)}{2h(\xi)}$.
			In this case $2^{1/\mu^*}$ can take the value $.833$.}
		\label{fig:ratio}\par\vspace{\intextsep}
		\includegraphics{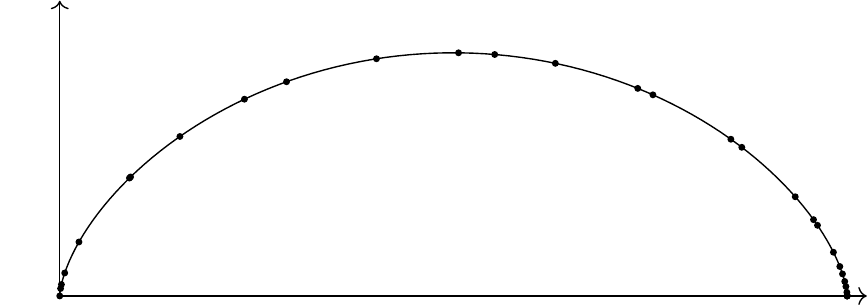}
		\caption{After five iterations, the expected value is already low
			enough that points start accumulating at the two ends.}
		\label{fig:h5}
	\end{figure}
	
	See Fig.~\ref{fig:h}, \ref{fig:ratio}, and \ref{fig:h5} for visualization.
	The punchline of this theorem
	is that its proof does not rely on any numerical result.
	As long as Formula~(\ref{eq:hmu}) holds
	for some choice of $h$ and $\mu^*$ the upper bound $\mu<\mu^*$ holds.
	As a corollary, \cite[Theorem~2]{MHU16} concludes
	$\mu\le3.639$.
	
	After that, \cite[Abstract]{FV14}
%	\fermat{This is a rather complicated one.
%		Urbanke dose not prove that $\mu\ge3.627$.
%		But Vardy says that Urbanke proves it.
%		But since Vardy claims that
%		all kernels of size $2$--$7$ have $\mu\ge3.627$,
%		it is a consequence that $\mu\ge3.627$ for \Arikans kernel.
%	}
	helps verify the estimate $\mu=3.627$.
	They also do an exhaustive computation to find
	larger kernels with better scaling exponent.
	An $8$-by-$8$ matrix \cite[below Table~1]{FV14} achieves $3.577$.
	A $16$-by-$16$ matrix \cite[above Section~VI]{FV14} achieves $3.356$.
	
	For general $q$-ary input channels,
	\cite{PU16} considers Reed--Solomon kernels and proves that
	they achieve scaling exponents arbitrary close to $2$
	as the field size (and thus the kernel size) grows.
	Being $2$ is optimal as has been shown in
	\cite{Dobrushin61}, \cite{Strassen64}, \cite{TZ00},
	\cite{Montanari01}, \cite{Hayashi09}, and \cite{PPV10}.
%	\fermat{There are like 4--5 papers each address different scenario
%		or with exotic notations.
%		I will try to read them and complete this part.}
	
	\cite{Hassani13} then conjectures, and provides strong evidence, that
	large kernels over the binary alphabet might suffice to achieve~$2$.
	It is then confirmed by \cite{FHMV17} by considering random binary kernels
	and $h(\xi)\coloneqq\(\xi(1-\xi)\)^\alpha$ for $\alpha$ close to~$0$.
	
	See Appendix~\ref{app:muruler} for comparison.
	\nocite{MHU15}

\subsubsection{Moderate Deviations Regime} \label{sec:mdr}

	In the previous two regimes,
%	\fermat{Is this the correct hyphenation?
%		re-gime?}
	either $I(W)-R_n$ or $P_n$ tends to zero
	while the other is (only) bounded from above.
	On the other hand, we want both of them to approach zero
	and to control the rate of convergence.
%	\fermat{Grammar?}
	See Fig.~\ref{fig:gb4} for visualization.
	
	The complicating factor is, that classical polar codes achieve
	certain $\beta$ and $\mu$ separately
	\textbf{does not} imply that classical polar codes can 
	\textbf{achieve both at the same time}.
	In principle we have to balance our efforts between
	reducing gap to capacity and reducing block error probability.
	
	\cite[Theorem~1]{GX13} states that there exists a $\mu'$
	(probably much larger than $\mu$) such that $\beta'=.49$ and $\mu'$ are
	achievable at the same time.
	We comment that this result sacrifices $\mu'$ to achieve
	a pretty good $\beta'$, just $0.01$ away from the best possibility.
	
	\cite{MHU16} introduces a ``interpolation'' result.
	\begin{theorem}\cite[Theorem~7 and Formula~(49)]{MHU16}
		\label{thm:gamma}
		Assume the $h$ and $\mu^*$ in Theorem~\ref{thm:hmu}.
		Let $\gamma$ be a free parameter such that
%		\fermat{Is this worth displaymath?}
		\[\frac1{1+\mu^*}<\gamma<1.\]
		Then 
		\[\beta'\coloneqq
			\gamma H_2^{-1}\(\frac{\gamma(\mu^*+1)-1}{\gamma\mu^*}\)
			\text{ and }\mu'\coloneqq\frac{\mu^*}{1-\gamma} \label{eq:gamma}\]
		are achievable at the same time.
		Here $H_2$ is the binary entropy function.
%		\fermat{Not the second homology group.}
	\end{theorem}
	
	When $\gamma\to1$ this recovers the error exponent $\beta'=\beta=1/2$.
	When $\gamma\to1/(1+\mu^*)$ this recovers a weaker
	scaling exponent $\mu'=1+\mu^*$.
	Our result, Theorem~\ref{thm:main}, recovers the true scaling exponent
	by Corollary~\ref{cor:recover}.
%	\fermat{Be careful $\mu^*$ and $\mu$.}
	
	\cite[Definition~1.1--1.4]{BGNRS18} proposed weaker notions
	to control $I(W)-R_n$ and $P_n$ 
	(where $P_n$ is exponential in $n$, instead of doubly exponential).
	That said, they derive some results based on much weaker assumptions
	\cite[Definition~1.5 and Theorem~1.6]{BGNRS18}.
	
	See Appendix~\ref{app:plane} for comparison.

\subsection{The Log-loglog Plot of the Path to Capacity}

	\def\G{-\log_2(\text{Gap})}
	\def\b{\log_2\(-\log_2(\text{bit error})\)}
	\def\B{\log_2\(-\log_2(\text{Block error})\)}
	\def\axes{
		\path(-.4,-.1);
		\draw[->](0,0)--(0,1)node[o,bly]{$-\log_2(\text{Gap})$};
		\draw[->](0,0)--(1,0)node[o,blx]{$\log_2\(-\log_2(P_n)\)$};
	}
	\begin{figure}\centering
		\tikz[gb]{
			\begin{scope}
				\clip(0,0)rectangle(1,1);
				\shade[inner color=darkgray](1,1)circle(.5);
			\end{scope}
			\axes
			\draw(.1,.1)•node[o,above]{$\bbB_1$}
				--(.2,.2)•node[o,below right]{$\bbB_2$}
				foreach\n in{3,...,9}{--(.\n-rand/20,.\n-rand/20)•}
				coordinate(B9);
			\draw[dotted](B9)--(1,1);
		}\caption{$\G$ ranges from $-\log(\text{Cpacity})$
			(which is at least one) to $+\infty$.
			The larger the better.
			And $\B$ ranges from $-\infty$ to $+\infty$.
			The larger the better.}
		\label{fig:log1}\par\vspace{\intextsep}
		\tikz[gb]{
			\shade[left color=white,right color=gray](.4,.2)rectangle(1,1);
			\draw(0,.2)node[o,left]{$-\log_2\(\text{Gap}\lb\)$}
				+(-.02,0)--+(1,0);
			\axes
			\draw(.1,.3)•node[o,above]{$\bbB_1$}
				--(.2,.3)•node[o,above]{$\bbB_2$}
				foreach\n in{3,...,9}{--(.\n-rand/20,.3-rand/20)•}
				coordinate(B9);
			\draw[dotted](B9)--(1,.3);
		}\caption{Error exponent regime:
			How fast we can move rightward while not moving too much downward.}
		\label{fig:log2}\par\vspace{\intextsep}
		\tikz[gb]{
			\shade[left color=white,right color=gray](.4,.2)rectangle(1,1);
			\draw foreach\i in{2,...,9}{(.\i,.2)--(.\i,1)};
			\draw(0,.2)node[o,left]{$-\log_2\(\text{Gap}\lb\)$}
				+(-.02,0)--+(1,0);
			\axes
		}\caption{Error exponent $\beta=1/2$ means that asymptotically
			each step is moving rightward by about $1/2$ units.}
		\label{fig:log3}\par\vspace{\intextsep}
		\tikz[gb]{
			\shade[bottom color=white,top color=gray](.2,.4)rectangle(1,1);
			\draw(.2,0)node[o,below,xshift=-9]
				{$\log_2\(-\log_2P\ub\)$}+(0,-.02)--+(0,1);
			\axes
			\draw(.3,.1)•node[o,right]{$\bbB_1$}
				--(.3,.2)•node[o,right]{$\bbB_2$}
				foreach\n in{3,...,9}{--(.3-rand/20,.\n-rand/20)•}
				coordinate(B9);
			\draw[dotted](B9)--(.3,1);
		}\caption{Scaling exponent regime:
			How fast we can move upward while not moving too much leftward.}
		\label{fig:log4}\par\vspace{\intextsep}
		\tikz[gb]{
			\shade[bottom color=white,top color=gray](.2,.4)rectangle(1,1);
			\draw foreach\i in{2,...,9}{(.2,.\i)--(1,.\i)};
			\draw(.2,0)node[o,below,xshift=-9]{$\log_2\(-\log_2P\ub\)$}
				+(0,-.02)--+(0,1);
			\axes
		}\caption{Scaling exponent $\mu=3.627$ means that asymptotically
			each step is moving upward by about $1/3.627$ units.}
		\label{fig:log5}
	\end{figure}
	\begin{figure}\centering
		\tikz[gb]{
			\shade[bottom color=white,top color=gray,shading angle=-45]
				(.2,.2)rectangle(1,1);
			\axes
			\draw(.1,.1)•node[o,above]{$\bbB_1$}
				--(.2,.2)•node[o,below right]{$\bbB_2$}
				foreach\n in{3,...,9}{--(.\n-rand/20,.\n-rand/20)•}
				coordinate(B9);
			\draw[dotted](B9)--(1,1);
		}\caption{Moderate deviations regime:
			How fast we can move in the direction up-right.}
		\label{fig:log6}\par\vspace{\intextsep}
		\tikz[gb]{
			\begin{scope}
				\clip(.2,.2)rectangle(1,1);
				\shade[bottom color=white,top color=gray,shading angle=-45]
					(.2,.2)rectangle(1,1);
				\draw foreach\i in{1,...,20}{(.2,.2+\i/10)--(.2+\i/10,.2)};
			\end{scope}
			\axes
		}\caption{Moderate deviations concern
			the joint performance of the previous two regimes.
			It should recover the previous two regimes as special cases.}
		\label{fig:log7}\par\vspace{\intextsep}
		\tikz[gb]{
			\begin{scope}
				\clip(.2,.2)rectangle(1,1);
				\draw foreach\i in{1,...,20}{
					(.2,.2+\i/11)to[bend right=10](.2+\i/10,.2)};
			\end{scope}
			\draw foreach\i in{3,...,9}{(.2,.\i)--+(-.1,0)};
			\draw foreach\i in{3,...,9}{(.\i,.2)--+(0,-.1)};
			\axes
		}\caption{Vertical segments $x=n\beta$ are
			those in Fig.~\ref{fig:log3} but clipped.
			Horizontal segments $y=n/\mu$ are
			those in Fig.~\ref{fig:log5} but clipped.
			Curves trace points $(\beta'n,n/\mu')$ as $\gamma$ varies.
			The curves match vertical segments but
			do not match horizontal ones.}
		\label{fig:log8}
	\end{figure}
	
	We have seen that in the context of polar coding
	there exist polar codes $\bbB_1,\bbB_2,\dotsc$
	such that the gap to capacity $I(W)-R_n$
	shrinks polynomially in $N_n=2^n$ as $n\to\infty$.
	Thus it is appropriate to compare $n$ to $-\log_2(\text{Gap to capacity})$,
	and the best ratio is the scaling exponent $\mu$.
	
	Similarly, there are $\bbB_1,\bbB_2,\dotsc$
	such that the block error probability $P_n$
	is as small as $2^{-N_n^{\text{certain fractional power}}}$.
	So it is appropriate to compare
	$\log_2\(-\log_2(\text{Block error probability})\)$ to $n$,
	and the best ratio is what we called error exponent%
	\footnote{We emphasize again that
		this is not the usual definition of the error exponent.}
	$\beta$
	
	Notice that the maximum of the bit error probabilities and
	the block error probability differ only by a factor of $N_n=2^n$.
	Thus $\b$ and $\B$ are of the same magnitude
	and we will use them interchangeably.
%	\fermat{Consider moving this paragraph down.}
	
	Consider locating $\bbB_1,\bbB_2,\dotsc$
	on the $\G$-versus-$\B$ plane.
	Then the goal of coding theory
	is such that those points converge to $(+\infty,+\infty)$,
	i.e.,~$\bbB_n$ ``moves'' in the direction of up-right.
	Or, think of a coding theorist standing at where $\bbB_n$ is
	only to construct $\bbB_{n+1}$ and jump to where $\bbB_{n+1}$ is.
	The one and only question is: how fast can we move
	in exchange for larger block length $N_n=2^n$?
	See Fig.~\ref{fig:log1} for visualization.
%	\fermat{.......}
	
	The error exponent measures
	how fast we can move rightward while not moving too much downward.
	The error exponent $\beta=1/2$ means asymptotically
	$\bbB_{n+1}$ is $1/2$ units to the right of $\bbB_n$.
	See Fig.~\ref{fig:log2} and \ref{fig:log3} for visualization.
	
	The scaling exponent measures
	how fast we can move upward while not moving too much leftward.
	The scaling exponent $\mu=3.627$ means asymptotically
	$\bbB_{n+1}$ is $1/3.627$ unit to the top of $\bbB_n$.
%	\fermat{or ``to the up''?}
	See Fig.~\ref{fig:log4} and \ref{fig:log5} for visualization.
	
	The moderate deviation regimes concerns
	the joint performance of the previous two regimes.
	Ideally it should take a free parameter $\gamma\in[0,1]$ such that
	\begin{easylist}[itemize]
		@ $\gamma$ controls the ``slope'' of the path $\bbB_n$.
		@ when $\gamma\to1$ the path $\bbB_n$ goes primarily rightward
			and recovers the error exponent regime;
		@ when $\gamma\to0$ the path $\bbB_n$ goes primarily upward
			and recovers the scaling exponent regime.
	\end{easylist}
	\smallskip
	See Fig.~\ref{fig:log6} and \ref{fig:log7} for visualization.
	
	Apart from the ideal case, we have seen that Theorem~\ref{thm:gamma},
	i.e.,~\cite[Theorem~7 and Formula~(49)]{MHU16},
	\begin{easylist}[itemize]
		@ recovers the error exponent $\beta=1/2$ when $\gamma\to1$;
		@ recovers a weaker scaling exponent $\mu^*+1>4.627$ when
			$\gamma\to1/(1+\mu^*)$.
	\end{easylist}
	\smallskip
	If we accept the Scaling Assumption \cite[Formula~(12)]{FV14}
	and the consequence that $\mu=3.627$,
	then there exists $h$ in the sense of Theorem~\ref{thm:hmu}
	such that $\mu^*$ is arbitrary close to $\mu=3.627$.
	Thus the suboptimality of Theorem~\ref{thm:gamma} is not that $\mu^*\neq\mu$
	but that at best we can only achieve $\mu'=1+\mu$, not $\mu'=\mu$.
	
	As a result, if we plot $\beta n$ and $n/\mu$ and trace the points
	$(\beta'n,n/\mu')$, there will be a discrepancy on the left hand side.
	See Fig.~\ref{fig:log8} for visualization.
	See Appendix~\ref{app:plane} for a more accurate plot.
	
	We will improve Theorem~\ref{thm:gamma} in Section~\ref{sec:main}.
	Before that,
	we brief the idea of Theorem~\ref{thm:gamma} in the next subsection.

\subsection{General Moving Strategy behind Theorem~\ref{thm:gamma}}
	
	\def\axes{
		\draw[->](0,0)--(0,1)node[o,bly]{$\G$};
		\draw[->](0,0)--(1,0)node[o,blx]{$\log_2\(-\log_2(P_n)\)$};
	}
	\begin{figure}\centering
		\tikz[gb]{
			\draw(.2,.2)•foreach\i in{1,...,6}{--++(.1,-.02)•}coordinate(X);
			\draw[->!](X)--+(.1,-.02)•;
			\axes
		}\caption{Moving rightward will cause moving downward a little bit.
			The longer the step length the harsher the penalty.
			The more the steps the softer the penalty.}
		\label{fig:pen1}\par\vspace{\intextsep}
		\tikz[gb]{
			\draw(.2,.2)•foreach\i in{1,...,7}{--++(.1,-.02)•}coordinate(X);
			\draw[->!](X)--(.2,.4)•;
			\axes
		}\caption{Moving upward will ``reset'' the $x$-coordinate.}
		\label{fig:pen2}
	\end{figure}
	\begin{figure}\centering
		\tikz[gb]{
			\draw(.2,.2)•foreach\i in{1,...,6}{--++(0,.1)•}
				foreach\i in{1,...,6}{--++(.1,-.02)•}coordinate(X);
			\draw[->!](X)--+(.1,-.02)•;
			\axes
		}\caption{The only productive arrangement seems to be
			to move upward and then move rightward.
			There is no way we can move rightward and then upward.
			Not to mention zigzagging.
			See Fig.~\ref{fig:pen4} for what actually happens.}
		\label{fig:pen3}\par\vspace{\intextsep}
		\tikz[gb]{
			\draw(.2,.2)•foreach\i in{1,...,6}{--++(0,.1)•}
				foreach\i in{1,...,6}{--++(-.02,0)•}coordinate(X);
			\draw[->!](X)--+(.8,-.1)•;
			\axes
		}\caption{What actually happens is that:
			recruit phase moves upward;
			then train phases moves slightly leftward;
			and finally retain phase moves rightward.}
		\label{fig:pen4}
	\end{figure}
	
	Granted to move $n$ steps,
	we may choose a free parameter $\gamma\in[0,1]$ and
	\begin{easylist}[itemize]
		@ move upward $n_0\coloneqq(1-\gamma)n$ steps to approach
			the $y$-coordinate $n_0/\mu$;
		@ move rightward $n_1\coloneqq\gamma n$ steps to approach
			the $x$-coordinate $\beta n_1$.
	\end{easylist}
	\smallskip
	However, just because we can reach $(x,y)=\(0,n_0/\mu\)$
	and $(x,y)=\(\beta n_1,0\)$ separately
	does not mean we can approach $(x,y)=\(\beta n_1,n_0/\mu\)$.
	Moving does not follow vector addition
	because it comes with some intrinsic penalties:
	\begin{easylist}[itemize]
		@ Moving rightward will cause moving downward a little bit.
			That is, to avoid error we discard bad synthetic channels,
			and that punishes the gap to capacity.
			(Fig.~\ref{fig:pen1}.)
		@ Moving upward will ``reset'' the $x$-coordinate.
			That is, to reduce the gap we collect more synthetic channels
			but cannot control their error probabilities.
%			\fermat{make it longer}
			(Fig.~\ref{fig:pen2}.)
	\end{easylist}
	\smallskip
	Therefore the only productive arrangement seems to be
	to move upward and then move rightward. (Fig.~\ref{fig:pen3}.)
%	\fermat{It is possible to introduce randomness in these figures.
%		Appropriate?}
	We now detail how to move and the cause of the penalties
	in the next subsection.
	We will demonstrate how to bypass these penalties in
	Section~\ref{sec:2pocket} and Section~\ref{sec:3pocket}.

\subsection{Detailed Movement: A Recruit-Train-Retain Model}

	\begin{figure}\centering
		\tikz[tree]{\setuptreecs
			\draw[hl=3](3,4)--(3,8);
			\draw(0,1)node(0W1){$\W11$};
			\foreach\i in{1,...,5}{
			\pgfmathtruncatemacro\I{2^\i}
				\foreach\j in{1,...,\I}{
					\pgfmathtruncatemacro\ii{\i-1}
					\pgfmathtruncatemacro\jj{(\j+1.1)/2}
					\draw(\i,\j)
						node[scale=1.44^(-\i+1)](\i W\j){$\W{\I}{\j}$}
						[line width=0.4*1.44^(-\i+1)](\i W\j)--(\ii W\jj);
				}
			}
		}\caption{The tree of synthetic channels
			and the collected $\W{2^{n_0}}j$ in the recruit phase.
			In general, they are not necessary consecutive.}
		\label{fig:tree1}\par\vspace{\intextsep}
		\tikz[tree]{\setuptreecs
			\draw[hl=5](5,13)--(5,16)(5,17)--(5,20)
				(5,21)--(5,24)(5,25)--(5,28)(5,29)--(5,32);
			\draw(0,1)node(0W1){$\W11$};
			\foreach\i in{1,...,5}{
			\pgfmathtruncatemacro\I{2^\i}
				\foreach\j in{1,...,\I}{
					\pgfmathtruncatemacro\ii{\i-1}
					\pgfmathtruncatemacro\jj{(\j+1.1)/2}
					\draw(\i,\j)
						node[scale=1.44^(-\i+1)](\i W\j){$\W{\I}{\j}$}
						[line width=0.4*1.44^(-\i+1)](\i W\j)--(\ii W\jj);
				}
			}
		}\caption{The descendants $\W{2^n}{2^{n_1}j-k}$
			of $\W{2^{n_0}}j$ in the train phase.}
		\label{fig:tree2}\par\vspace{\intextsep}
		\tikz[tree]{\setuptreecs
			\draw[hl=5](5,14)--(5,16)(5,18)--(5,20)
				(5,22)--(5,24)(5,26)--(5,28)(5,30)--(5,32);
			\draw(0,1)node(0W1){$\W11$};
			\foreach\i in{1,...,5}{
			\pgfmathtruncatemacro\I{2^\i}
				\foreach\j in{1,...,\I}{
					\pgfmathtruncatemacro\ii{\i-1}
					\pgfmathtruncatemacro\jj{(\j+1.1)/2}
					\draw(\i,\j)
						node[scale=1.44^(-\i+1)](\i W\j){$\W{\I}{\j}$}
						[line width=0.4*1.44^(-\i+1)](\i W\j)--(\ii W\jj);
				}
			}
		}\caption{The retained $\W{2^n}{2^{n_1}j-k}$ in the retain phase.}
		\label{fig:tree3}
	\end{figure}

%	\fermat{I am not sure if ``Recruit-Train-Retain'' is a good name.
%		If not, we can call it ``the $\Gamma$'' Construction}
	Moving upward $n_0\coloneqq(1-\gamma)n$ steps is straightforward.
%	\fermat{\cite{MHU16} uses the notation $\gamma,n_0,n_1$.}

\subsubsection{Recruit Phase}

	Set a goal $P\ub$ and collect as many synthetic channels $\W{2^{n_0}}j$
%	\fermat{This is the standard notation introduced in \cite{Arikan09}}
	as possible such that the error probability does not exceed $P\ub$.
	Notice that the maximum and the sum of bit error probabilities
	differ only by a negligible factor of $2^{n_0}<2^n$
	so we do not distinguish which one we are talking about.
	See Fig.~\ref{fig:tree1}.
	
	Setting such a goal $P\ub$
	will pin us at the $x$-coordinate $\log(-\log P\ub)$.
	By Theorem~\ref{thm:hmu} or the estimate that $\mu=3.627$
	\cite[Abstract]{FV14} we will collect so many synthetic channels
	such that the gap to capacity is $O\(2^{n_0/\mu}\)$.
	This will bring us to the $y$-coordinate $n_0/\mu$.
	
	With these $\W{2^{n_0}}j$ in our pocket,
	moving rightward $n_1\coloneqq\gamma n$ steps consists of two phases.

\subsubsection{Train Phase}

	For each $\W{2^{n_0}}j$ in our pocket, remove it and put
	both $\W{2^{n_0+1}}{2j-1}$ and $\W{2^{n_0+1}}{2j}$ in our pocket.
	Doing so will maintain the gap to capacity
	and double the error probability.
	Thus we are actually moving leftward, but not too much.
	Repeat this doubling process $n_1$ times.
	Each $\W{2^{n_0}}j$ leads to $2^{n_1}$ descendants
	of the form $\W{2^n}{2^{n_1}j-k}$ for $k=0,\dotsc,2^{n_1}-1$.
	See Fig.~\ref{fig:tree2}.
%	\fermat{\Arikan uses the word ``give birth'' and ``children''}

\subsubsection{Retain Phase}

	For each $\W{2^n}{2^{n_1}j-k}$ in our pocket generated by $\W{2^{n_0}}j$,
	its error probability is
	doubled $\weight(k)$ times and squared $n_1-\weight(k)$ times.
	Here $\weight(k)$ is the Hamming weight of $k$ written in binary.
	
	If $P\ub$ is small enough the order of doubling and squaring is minor.
	What matters is the total number of squaring.
	We therefore set a threshold $\epsilon$ and discard those
	whose error probability is squared less than $\epsilon n_1$ times.
	This will bring us to the $x$-coordinate $\epsilon n_1$.
	See Fig.~\ref{fig:tree3}.
	
	Here comes the penalty: Discarding synthetic channels in our pocket
	increases the gap to capacity.
	For instance if $\epsilon\ge1/2$ then
	half of synthetic channels in our pocket are unqualified.
	Even if $\epsilon<1/2$, the portion of unqualified synthetic channels
	are asymptotically $2^{-n_1\(1-H_2(\epsilon)\)}$
	for $H_2$ the binary entropy function.
	If $n_1\(1-H_2(\epsilon)\)<n_0/\mu$,
	we are forced to return to the $y$-coordinate
	$n_1\(1-H_2(\epsilon)\)$ from $n_0/\mu$.
	
	When $\gamma\to1$, we have a lot of quota of moving rightward
	and interestingly the effect of moving downward is diluted and negligible.
	(Casually speaking, training a lot increases the retention rate.)
%	\fermat{Consider remove this paragraph.}
	We will see in the next subsection
	the obstacle to $\mu'\to\mu$ when $\gamma\to0$.

\subsection{The Main Obstacle to \texorpdfstring{$\mu'\to\mu$}{μ'→μ}}
	\label{sec:obstacle}

	\begin{figure}\centering
		\tikz[gb]{
			\draw(.2,.2)•foreach\i in{1,...,6}{--++(0,.1)•}
				foreach\i in{1,...,4}{--++(-.02,0)•}coordinate(X);
			\draw[->!](X)--+(.3,-.4)•;
			\axes
		}\caption{Sometimes moving rightward comes with significant penalty.}
		\label{fig:0.9}\par\vspace{\intextsep}
		\tikz[gb]{
			\draw(.2,.2)•foreach\i in{1,...,4}{--++(0,.1)•}
				foreach\i in{1,...,6}{--++(-.02,0)•}coordinate(X);
			\draw[->!](X)--+(.6,-.1)•;
			\axes
		}\caption{To avoid the penalty we better not to move so high.}
		\label{fig:0.61}
	\end{figure}
	
	For example let $\gamma=0.1$, so $n_0=0.9n$ and $n_1=0.1n$.
	\begin{easylist}[itemize]
		@ First move upward $0.9n$ steps.
			Now we are at the $y$-coordinate $0.9n/\mu\ge0.24n$.
		@ For each $\W{2^{0.9n}}j$ in our pocket,
			generate $2^{0.1n}$ descendants of the form
			$\W{2^n}{2^{0.1n}j-k}$ for $k=0,1,\dotsc,2^{0.1n}-1$.
		@ For these $2^{0.1n}$ descendants we have several choices:
			@@ Keep all of them.
				Then the error probability is doubled $0.1n$ steps,
				so we are actually moving leftward.
				No progress is made.
			@@ Keep all but one.
				Then all errors probabilities are squared at least once,
				so we are moving rightward by one unit.
				But this means we lose $2^{-0.1n}$ of synthetic channels.
				The code rate will drop by about $2^{-0.1n}$.
				So the gap to capacity is at least $2^{-0.1n}$.
				This will ``reset'' our $y$-coordinate to $0.1n$ from $0.24n$.
				See Fig.~\ref{fig:0.9} for visualization.
			@@ Discard more then one.
				Then we lose even more synthetic channels/rate/$y$-coordinate.
	\end{easylist}
	\smallskip
	In this particular case we should not go to $y=0.24n$ in the first place.
	An obviously better way is to stop at $y=0.1n$ after $0.39n$ steps,
	and then move rightward using $(1-0.39)n=0.61n$ steps.
	This will bring us to an even larger, better $x$-coordinate $0.17n$
	while maintaining a larger, better $y$-coordinate.
	See Fig.~\ref{fig:0.61} for visualization.
	See Appendix~\ref{app:plane} for comparison.

	By some trivial calculation one can show that
	if we are allowed to move rightward only $n/(1+\mu)$ steps
	then it is better not to redeem those moves at all.
	This explains why \cite[Theorem~7]{MHU16}
	recovers scaling exponent $1+\mu$ when $\gamma\to1/(1+\mu)$
	instead of $\mu$ for $\gamma\to0$.
	
	\begin{figure}\centering
		\tikz[tree]{\setuptreecs
			\draw[hl=3](3,4)--(3,8);
			\draw[hm=4](4,4)--(4,6);
			\draw(0,1)node(0W1){$\W11$};
			\foreach\i in{1,...,5}{
			\pgfmathtruncatemacro\I{2^\i}
				\foreach\j in{1,...,\I}{
					\pgfmathtruncatemacro\ii{\i-1}
					\pgfmathtruncatemacro\jj{(\j+1.1)/2}
					\draw(\i,\j)
						node[scale=1.44^(-\i+1)](\i W\j){$\W{\I}{\j}$}
						[line width=0.4*1.44^(-\i+1)](\i W\j)--(\ii W\jj);
				}
			}
		}\caption{The primary pocket (larger region)
			and the secondary pocket (smaller region) in the recruit phase.
			Notice that they do not ``overlap''.}
		\label{fig:tree4}\par\vspace{\intextsep}
		\tikz[tree]{\setuptreecs
			\draw[hl=5](5,13)--(5,16)(5,17)--(5,20)
				(5,21)--(5,24)(5,25)--(5,28)(5,29)--(5,32);
			\draw[hm=5](5,7)--(5,8)(5,9)--(5,10)(5,11)--(5,12);
			\draw(0,1)node(0W1){$\W11$};
			\foreach\i in{1,...,5}{
			\pgfmathtruncatemacro\I{2^\i}
				\foreach\j in{1,...,\I}{
					\pgfmathtruncatemacro\ii{\i-1}
					\pgfmathtruncatemacro\jj{(\j+1.1)/2}
					\draw(\i,\j)
						node[scale=1.44^(-\i+1)](\i W\j){$\W{\I}{\j}$}
						[line width=0.4*1.44^(-\i+1)](\i W\j)--(\ii W\jj);
				}
			}
		}\caption{The descendants in the train phase.
			Note that synthetic channels in the primary pocket
			get more chances to square their erasure probabilities.}
		\label{fig:tree5}\par\vspace{\intextsep}
		\tikz[tree]{\setuptreecs
			\draw[hl=5](5,14)--(5,16)(5,18)--(5,20)
				(5,22)--(5,24)(5,26)--(5,28)(5,30)--(5,32);
			\draw[hm=5](5,8)circle(0)(5,10)circle(0)(5,12)circle(0);
			\draw(0,1)node(0W1){$\W11$};
			\foreach\i in{1,...,5}{
			\pgfmathtruncatemacro\I{2^\i}
				\foreach\j in{1,...,\I}{
					\pgfmathtruncatemacro\ii{\i-1}
					\pgfmathtruncatemacro\jj{(\j+1.1)/2}
					\draw(\i,\j)
						node[scale=1.44^(-\i+1)](\i W\j){$\W{\I}{\j}$}
						[line width=0.4*1.44^(-\i+1)](\i W\j)--(\ii W\jj);
				}
			}
		}\caption{The retained in the retain phase.
			The policy ``to keep those who square once''
			is harsher in the secondary pocket.
			But the damage to the rate is ameliorated since
			the secondary pocket contains fewer to start with.}
		\label{fig:tree6}
	\end{figure}
	
	We will demonstrate how to bypass this obstacle in the next subsection,
	in Section~\ref{sec:3pocket}, and finally in Theorem~\ref{thm:main}.

\subsection{Bypassing Obstacle: Two-Pocket Recruit-Train-Retain}
	\label{sec:2pocket}

	\begin{figure}\centering
		\tikz[gb]{
			\draw(.2,.2)•foreach\i in{1,...,4}{--++(0,.1)•}coordinate(Y)
				foreach\i in{1,...,6}{--++(-.02,0)•}coordinate(X);
			\draw[->!](X)--+(.6,-.1)•;
			\draw(Y)•foreach\i in{5,...,6}{--++(-0,.1)•}
				foreach\i in{1,...,4}{--++(-.02,0)•}coordinate(X);
			\draw[->!](X)--+(.3,-.05)•;
			\axes
		}\caption{Reaching a higher $y$-coordinate using a secondary pocket
			while not risking loosing too much from the primary pocket.}
		\label{fig:2pocket}\par\vspace{\intextsep}
		\tikz[gb]{
			\draw(.2,.2)•foreach\i in{1,...,4}{--++(0,.1)•}coordinate(Y)
				foreach\i in{1,...,6}{--++(-.02,0)•}coordinate(X);
			\draw[->!](X)--+(.6,-.1)•;
			\draw(Y)•foreach\i in{5,...,5}{--++(-0,.1)•}coordinate(Y)
				foreach\i in{1,...,5}{--++(-.02,0)•}coordinate(X);
			\draw[->!](X)--+(.58,-.1)•;
			\draw(Y)•foreach\i in{6,...,6}{--++(-0,.1)•}
				foreach\i in{1,...,4}{--++(-.02,0)•}coordinate(X);
			\draw[->!](X)--+(.56,-.1)•;
			\axes
		}\caption{Reaching a higher $y$-coordinate using three pockets.}
		\label{fig:3pocket}
	\end{figure}

%	Say we are granted to move $n$ steps.
%	\fermat{we apply the RTR twice after dividing channels into two pockets.}
	We prepare two pockets to hold synthetic channels
	and apply the recruit-train-retain trick to both pockets separately.
	The advantage is that we can implement different policy in different pocket.
	
	\begin{easylist}[itemize]
		@ Collect in a primary pocket synthetic channels $\W{2^{0.7n}}j$ with
			low error probability, i.e.,~move upward $0.7n$ steps
			to approach the $y$-coordinate $0.7n/\mu>0.19n$.
			See Fig.~\ref{fig:tree4}.
		@ For each $\W{2^{0.7n}}j$ generate $\W{2^n}{2^{0.3n}j-k}$ and discard
			those whose error probability is squared less than $0.01n$ times.
			Notice $0.3n\(1-H_2(0.02/0.3)\)>0.19n$.
			We lose $2^{-0.19n}$ of synthetic channels in the primary pocket,
			which is satisfactorily few.
			Visually, we move rightward $0.3n$ steps to approach
			the $x$-coordinate $0.02n$ while maintaining the $y$-coordinate
			$0.19n$.
			See Fig.~\ref{fig:tree5} and \ref{fig:tree6}.
		@ At the same time, collect in a secondary pocket synthetic channels
			$\W{2^{0.9n}}l$ that are not a descendant of a $\W{2^{0.7n}}j$
			in the primary pocket.
			This secondary pocket should contain at most $2^{-0.19n}$
			(the gap of the primary pocket) of synthetic channels.
			That is, a very thin branch approaches
			the $y$-coordinate $0.9n/\mu>0.24n$.
			See Fig.~\ref{fig:tree4}.
		@ For each $\W{2^{0.9n}}l$ generate $\W{2^n}{2^{0.1n}l-m}$ and discard
			those whose error probability is squared less than $0.01n$ times.
			Notice $0.1n\(1-H_2(0.01/0.1)\)>0.05n$.
			We lose $2^{-0.05n}$ of synthetic channels in the secondary pocket,
			approximately $2^{-0.19n-0.05n}=2^{-0.24n}$
			of all synthetic channels, satisfactorily few.
			See Fig.~\ref{fig:tree5} and \ref{fig:tree6}.
		@ Overall, we reach the $y$-coordinate $0.24n$
			and the $x$-coordinate $0.01n$.
	\end{easylist}
	\smallskip
	See Fig.~\ref{fig:2pocket} for visualization.
	This already surpasses Theorem~\ref{thm:gamma}.
	See Appendix~\ref{app:plane} for comparison.
	
	In Section~\ref{sec:obstacle} we see the conflict between
	retaining synthetic channels to maintain the gap to capacity
	and discarding bad performance ones to reduce the error probability.
	From the example above we see that by dividing  synthetic channels into
	two pockets, each pocket may have its own retain-discard policy.
	This dissolves the conflict.
	
	And we can do better.
	In Theorem~\ref{thm:main}, we will declare a large number of pockets
	to minimize the conflict.
	Before that, we demonstrate a three-pocket trick in the next subsection.

\subsection{One More Example: Three-Pocket Recruit-Train-Retain}
	\label{sec:3pocket}

	Say we are granted to move $n$ steps.
	\begin{easylist}[itemize]
		@ Collect in a primary pocket synthetic channels $\W{2^{0.7n}}j$ with
			low error probability, i.e.,~move upward $0.7n$ steps
			to approach the $y$-coordinate $0.7n/\mu>0.19n$.
		@ For each $\W{2^{0.7n}}j$ generate $\W{2^n}{2^{0.3n}j-k}$ and discard
			those whose error probability is squared less than $0.01n$ times.
			Notice $0.3n\(1-H_2(0.02/0.3)\)>0.19n$.
			We lose $2^{-0.19n}$ of synthetic channels in the primary pocket,
			which is satisfactorily few.
			Visually, we move rightward $0.3n$ steps to approach
			the $x$-coordinate $0.02n$ while maintaining the $y$-coordinate
			$0.19n$.
		@ At the same time, collect in a secondary pocket synthetic channels
			$\W{2^{0.8n}}l$ that are not a descendant
			of some $\W{2^{0.7n}}j$ in the primary pocket.
			This secondary pocket should contain at most $2^{-0.19n}$
			(the gap of the primary pocket) of synthetic channels.
			That is, a very thin branch approaches
			the $y$-coordinate $0.8n/\mu>0.22n$.
		@ For each $\W{2^{0.8n}}l$ generate $\W{2^n}{2^{0.2n}l-m}$ and discard
			those whose error probability is squared less than $0.02n$ times.
			Notice $0.2n\(1-H_2(0.02/0.2)\)>0.10n$.
			We lose $2^{-0.10n}$ of synthetic channels in the secondary pocket,
			approximately $2^{-0.19n-0.10n}=2^{-0.29n}$
			of all synthetic channels, satisfactorily few.
		@ At the same time, collect in a tertiary pocket synthetic channels
			$\W{2^{0.9n}}p$ that are
			neither a descendant of some $\W{2^{0.7n}}j$ in the primary pocket 
			or a descendant of some $\W{2^{0.8n}}l$ in the secondary pocket.
			This tertiary pocket should contain at most $2^{-0.22n}$
			(the gap of the secondary pocket) of synthetic channels.
			That is, an even thinner branch approaches
			the $y$-coordinate $0.9n/\mu>0.24n$.
		@ For each $\W{2^{0.9n}}p$ generate $\W{2^n}{2^{0.1n}p-q}$ and discard
			those whose error probability is squared less than $0.02n$ times.
			Notice $0.1n\(1-H_2(0.02/0.1)\)>0.02n$.
			We lose $2^{-0.02n}$ of synthetic channels in the secondary pocket,
			approximately $2^{-0.22n-0.02n}=2^{-0.24n}$
			of all synthetic channels, satisfactorily few.
		@ Overall, we reach the $y$-coordinate $0.24n$
			and the $x$-coordinate $0.02n$.
	\end{easylist}
	\smallskip
	See Fig.~\ref{fig:3pocket} for visualization.
	This surpasses Theorem~\ref{thm:gamma} and
	the example in Section~\ref{sec:2pocket}.
	See Appendix~\ref{app:plane} for comparison.
	
	We now give a self-contained introduction of
	polar codes and related terminologies in the next section.

\section{Preliminary}

%	\fermat{TODO:
%		@ Treat polar coding as a bb.
%		@ Focus on \Bha process, not the decoder etc.
%		@ Introduce the transformation $(W,W)\mapsto(W',W'')$ as a bb.
%		@ Consider repeat Theorem~\ref{thm:hmu} again.
%		@ Introduce synthetic channel and ``descendant''.}

\subsection{Binary Erasure Channels}

	A \emph{binary erasure channel $W$ of erasure probability $Z(W)$}
	has input alphabet $\bbF_2$ and output alphabet $\bbF_2\cup\{?\}$.
	The properties of the channel are described by the probability mass function
	\begin{align}
		W(1|0)=W(0|1) &= 0; \\
		W(?|0)=W(?|1) &= Z(W); \\
		W(0|0)=W(1|1) &= 1-Z(W).
	\end{align}
	The capacity of this channel is $I(W)=1-Z(W)$.
%	\fermat{Put some references. Presumably van Lint}
%	We will only consider binary erasure channels in this work.

\subsection{Channel Polarization} \label{sec:butterfly}

	On binary erasure channels,
	channel polarization consists of the following pair of building blocks
	\[\tikz[BB]\node[E]{};\text{ and }\tikz[BB]\node[D]{};.\]
	This pair of building blocks has the ability that if we wrap-up
	a pair of i.i.d. channels $W$ like
	\[\tikz[BB]\setupbbcs\draw
		(-1,0)node[E](E){}(+1,0)node[D](D){}
		(E.NE)--node[o,above right]{$W$}(D.NW)
		(E.SE)--node[o,below left ]{$W$}(D.SW)
		(E.NW)node[left]{$A$}(D.NE)node[right]{$B$}
		(E.SW)node[left]{$C$}(D.SE)node[right]{$D$}
	;,\]
	then point $A$ to point $B$ forms a synthetic binary erasure channel $W'$
	with erasure probability $Z(W')=1-\(1-Z(W)\)^2$, while
	point $C$ to point $D$ forms another synthetic binary erasure channel $W''$
	with erasure probability $Z(W'')=Z(W)^2$.
%	\fermat{Double check if this is compatible to \Arikans notation}
	
	%Here comes the brilliant idea:
	A crucial, novel idea in the construction of polar codes is that
	we may begin with four i.i.d channels $W$ and wrap them up as
	\[\tikz[BB]\setupbbcs\draw
	(-2,1)node[F](E1){}(-1,1)node[E](E2){}(+1,1)node[D](D1){}(+2,1)node[F](D2){}
	(-2,0)node[F](E3){}(-1,0)node[E](E4){}(+1,0)node[D](D3){}(+2,0)node[F](D4){}
		(E1.NE)--(E2.NW)(E2.NE)--node[above right]{$W$}(D1.NW)(D1.NE)--(D2.NW)
		(E1.SE)--(E4.NW)(E2.SE)--node[below left ]{$W$}(D1.SW)(D1.SE)--(D4.NW)
		(E3.NE)--(E2.SW)(E4.NE)--node[above right]{$W$}(D3.NW)(D3.NE)--(D2.SW)
		(E3.SE)--(E4.SW)(E4.SE)--node[below left ]{$W$}(D3.SW)(D3.SE)--(D4.SW)
	;.\]
	This setup is equivalent to four synthetic channels
	\[\tikz[BB]\setupbbcs\draw
		(-2,1)node[F](E1){}(+2,1)node[F](D2){}
		(-2,0)node[F](E3){}(+2,0)node[F](D4){}
		(E1.NE)--node[above right]{$W'$}(D2.NW)
		(E1.SE)--node[below left ]{$W'$}(D2.SW)
		(E3.NE)--node[above right]{$W''$}(D4.NW)
		(E3.SE)--node[below left ]{$W''$}(D4.SW)
	;\]
	where the two occurrences of $W'$ are independent
	and the two occurrences of $W''$ are independent.
	Thus we can further wrap them
	\[\tikz[BB]\setupbbcs\draw
	(-2,1)node[E](E1){}(-1,1)node[E](E2){}(+1,1)node[D](D1){}(+2,1)node[D](D2){}
	(-2,0)node[E](E3){}(-1,0)node[E](E4){}(+1,0)node[D](D3){}(+2,0)node[D](D4){}
		(E1.NE)--(E2.NW)(E2.NE)--node[above right]{$W$}(D1.NW)(D1.NE)--(D2.NW)
		(E1.SE)--(E4.NW)(E2.SE)--node[below left ]{$W$}(D1.SW)(D1.SE)--(D4.NW)
		(E3.NE)--(E2.SW)(E4.NE)--node[above right]{$W$}(D3.NW)(D3.NE)--(D2.SW)
		(E3.SE)--(E4.SW)(E4.SE)--node[below left ]{$W$}(D3.SW)(D3.SE)--(D4.SW)
	;\]
	and obtain four synthetic channels $(W')'$, $(W')''$, $(W'')'$, $(W'')''$
	with erasure probabilities
	\begin{align}
		Z((W')') &= 1-(1-Z(W'))^2 = 1-(1-Z(W))^4; \\
		Z((W')'') &= Z(W')^2 = (1-(1-Z(W)^2))^2; \\
		Z((W'')') &= 1-(1-Z(W''))^2 = 1-(1-Z(W)^2)^2; \notag\\
		Z((W'')'') &= Z(W'')^2 = Z(W)^4.
	\end{align}
	
	The construction does not stop here.
	We may let $\W11\coloneqq W$ and inductively construct synthetic channels
	\[\W{2N}{2j-1}\coloneqq\(\W Nj\)',\qquad\W{2N}{2j}\coloneqq\(\W Nj\)''.\]
	We call $\W{MN}{Mj-k}$ a \emph{descendant} of $\W Nj$ if
	the former is obtained from the later in this way, i.e.,~$0\le k<M$.
	Conversely we call $\W Nj$ an \emph{ancestor} of $\W{MN}{Mj-k}$.
	
	See Fig.~\ref{fig:polar8} for a larger construction.
	See Appendix~\ref{app:butterfly} for an even larger construction.
	\begin{figure}
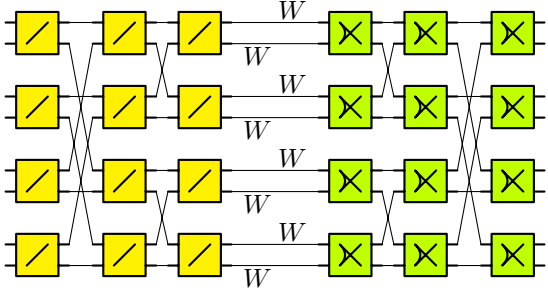
\centering
		\tikz[BB]{\setupbbcs
			\foreach\J[count=\j]in{00,01,10,11}{
				\draw foreach\l in{1,2,3}{
						(-\l,\j)node[E](\l E\J){}(+\l,\j)node[D](\l D\J){}}
					(1E\J.NE)--node[above right]{$W$}(1D\J.NW)
					(1E\J.SE)--node[below left ]{$W$}(1D\J.SW);}
			\def\DRAW#1#2{
				\draw foreach\j in{0,1}{
					(#1E\k1.NW)--(#2E\k1.NE)(#1D\k1.NE)--(#2D\k1.NW)
					(#1E\k1.SW)--(#2E\k0.NE)(#1D\k1.SE)--(#2D\k0.NW)
					(#1E\k0.NW)--(#2E\k1.SE)(#1D\k0.NE)--(#2D\k1.SW)
					(#1E\k0.SW)--(#2E\k0.SE)(#1D\k0.SE)--(#2D\k0.SW)};}
			\def\k#1{\j#1}\DRAW12
			\def\k#1{#1\j}\DRAW23
		}\caption{A larger polar code construction.
			It generates eight synthetic channels.
			From top to bottom:
			$((W')' )'$, $((W')' )''$, $((W'')' )'$, $((W'')' )''$,
			$((W')'')'$, $((W')'')''$, $((W'')'')'$, $((W'')'')''$.
			Or equivalently:
			$\W81$, $\W82$, $\W85$, $\W86$, $\W83$, $\W84$, $\W87$, $\W88$.}
		\label{fig:polar8}
	\end{figure}

\subsection{Apply Polar Coding in Communication}

	Choose an $N_n=2^n$ and among $N_n$ synthetic channels
	$\W{N_n}1,\W{N_n}2,\dotsc\W{N_n}{N_n}$
	choose a subset $\clA_n$ of synthetic channels.
	To communicate, send messages through synthetic channels in $\clA_n$
	and send predictable symbols (for instance, all zero)
	through synthetic channels not in $\clA_n$.
	A subset $\clA_n$ is understood as a \emph{polar code}.
	
	The \emph{block length} $N_n$ associated to this code,
	equivalently to $\clA_n$, is $N_n=2^n$.
	The associated \emph{code rate} $R_n$ is $|\clA_n|/N_n$.
	The associated \emph{block error probability} $P_n$ is the probability
	that any synthetic channel in $\clA_n$ erases the message.
	Clearly this quantity is less than
	the sum of all erasure probabilities $Z\(\W{2^n}j\)$
	of the synthetic channels $\W{2^n}j$ in $\clA_n$, by the union bound.
	
	On the one hand, the sum of erasure probabilities
	overestimates the block error probability $P_n$.
	On the other hand, the maximal erasure probability
	differ from the sum by a scaler of $N_n=2^n$.
	This becomes negligible once we take the logarithm twice,
	so we do not expect to gain from a more precise estimate.
	For soundness, however, we will argue only with the sum, not the maximum.
	(Nevertheless, for the tightness of the union bound, see \cite{BPT13}.)
	
	The goal of this work is to understand the relation among
	block length $N_n$, code rate $R_n$, and the block error probability $P_n$
	(bounded from above by the sum of erasure probabilities
	of synthetic channels in $\clA_n$),
	using terminologies defined in the next subsection.

\subsection{Error Exponent and Scaling Exponent}

	Let $\clA_n$ be a series of polar codes with block length $N_n=2^n$,
	code rate $R_n$, and block error probability $P_n$.
	The (equivalent) \emph{error exponent} of this series of codes is%
%	\fermat{Either call it ``equivalent error exponent''
%		or ``empirical error exponent'' or ``instance error exponent''.}
	\footnote{We emphasize the third time that
		this is not the usual definition of error exponent.}
	\[\beta'\coloneqq\liminf_{n\to\infty}\frac{\log\(-\log P_n\)}{\log N_n}.\]
	The \emph{error exponent of polar coding} $\beta$ is the supremum of
	(equivalent) error exponents taken over all series of polar codes.
	See Section~\ref{sec:eer} for previous works.
	
	The (equivalent) \emph{scaling exponent} of this series of codes is
	\[\mu'\coloneqq\limsup_{n\to\infty}\frac{-\log(N_n)}{\log\(I(W)-R_n\)}.\]
	The \emph{scaling exponent of polar coding} $\mu$ is the infimum of
	(equivalent) scaling exponents taken over all series of polar codes.
	See Section~\ref{sec:ser} for previous works.
	
	The goal of this work is to understand what pair of
	$(\beta',\mu')$ is achievable simultaneously by a series of polar codes.
	In general, our solution is a trade-off between $\beta'$ and $\mu'$.
%	\fermat{fixit.}
	See Section~\ref{sec:mdr} for previous works.

\subsection{\Bha Process}

	To describe the erasure probabilities of synthetic channels better,
	define a discrete Markov process by
	letting $Z_0\coloneqq Z(W)$ and inductively
	\[Z_{n+1}\coloneqq\begin{cases*}
		1-\(1-Z_n\)^2 & with probability $1/2$; \\
		Z_n^2 & with probability $1/2$.
	\end{cases*}\]
	This is called the \emph{\Bha process}.
	
	In other words, $Z_n$ is the erasure probability $Z\(\W{2^n}{j_n}\)$
	of a uniformly randomly chosen synthetic channel $\W{2^n}{j_n}$
	such that $j_n=2j_{n-1}-1$ or $j_n=2j_{n-1}$.
	Making it a process simplifies some notation.
	For example, the fact that $1-\(1-Z_n\)^2+Z_n^2=2Z_n$
	is equivalent to $Z_n$ being martingale.
	Consequently $\bbE\left[Z_n\right]=Z_0=Z(W)=1-I(W)$.
	
	We now quote some lemmata from previous works
	to illustrate how $Z_n$ works in the next subsection.

\subsection{Lemmata From/Inspired by Previous Works}

	See also \cite[Formula~(11)]{MHU16} for the definition of
	\Bha process $Z_n$.
%	\fermat{This is the second time we use \Bha process.
%		This reference is different from the previous one.}
	
	\begin{lemma}\cite[Lemma~6 and Formula~(29)]{MHU16} \label{lem:hrho1}
		Let $h:[0,1]\to[0,1]$
		be such that $h(0)=h(1)=0$ and $h(\xi)>0$ otherwise.
		Assume
		\[\sup_{0<\xi<1}\frac{h\(\xi^2\)+h\(2\xi-\xi^2\)}{2h(\xi)}
			\le2^{-\rho_1} \label{eq:hrho1}\]
		for some $\rho_1\le1/2$.
		Fix an $\alpha\in(0,1)$,
		then for any $\delta\in(0,1)$ and $m\in\bbN$
		\[\bbE\left[\(Z_m(1-Z_m)\)^\alpha\right]\le
			\frac1\delta\(2^{-\rho_1}+\frac{\sqrt2c_3\delta}{1-\delta}\)^m
				\label{eq:delta}\]
		for some constant $c_3$
		depending on $h,\rho_1,\alpha$, but not $m,\delta$.
	\end{lemma}
	\begin{IEEEproof}
		Omitted.
	\end{IEEEproof}
	
	\begin{lemma}Inspired by \cite[Lemma~5 and Formula~(29)]{MHU16}.
		\label{lem:alpha}
		Fix an $\alpha\in(0,1)$, a $\rho\le1/2$, a $c_1>0$, and an $D>1$.
		Assume for all $m\in\bbN$
		\[\bbE\left[\(Z_m(1-Z_m)\)^\alpha\right]
			\le c_12^{-m\rho}.\]
		Then for all $m\in\bbN$
		\[\bbP\(Z_m\le P\ub2^{-Dm}\)\ge I(W)-c_22^{-m(\rho-D\alpha)}\]
		for some $c_2$ depending on $P\ub,\alpha,c_1,m$, but not $m$.
	\end{lemma}
	\begin{IEEEproof}
		See Appendix~\ref{pf:alpha}.
	\end{IEEEproof}
	
	\begin{lemma} \label{lem:D}
		Assume the $h,\mu^*$ in Theorem~\ref{thm:hmu}.
		Then for any fixed $D>1$,
		\[\bbP\(Z_m\le P\ub2^{-Dm}\)\ge I(W)-O\(2^{-m/\mu^*}\)\]
		as $m$ varies.
	\end{lemma}
	\begin{IEEEproof}
		See Appendix~\ref{pf:D}.
	\end{IEEEproof}

\section{Main Result} \label{sec:main}

	\begin{theorem} \label{thm:main}
%		\fermat{Should we emphasis that this is our result?
%			For example put above ``We propose the following criterion.''}
		Assume the $h$ and $\mu^*$ in Theorem~\ref{thm:hmu}.
		If for all $\pi\in[0,1]$
%		\fermat{Seeking for better parameterization.}
		\[\frac{1-\pi}{\mu'-\mu^*\pi}+H_2\(\frac{\beta'\mu'}{\mu'-\mu^*\pi}\)
			<1 \label{eq:prere}\]
		then $(\beta',\mu')$ is achievable.
		More Precisely, for $n$ large enough there exists a polar code $\bbB_n$
		of blocklength $2^n$
%		\fermat{Do I emphasis that
%			their block lengths $2^1,2^2,\dotsc$ are consecutive.
%			That is, we do not have to skip any possible block length.}
		such that
%		\fermat{The $O$ symbol in gap can be taken off.
%			Just shrink $\mu'$ a little bit.}
		\[P_n\le 2^n\cdot2^{-2^{\beta'n}};\qquad
			I(W)-R_n=O\(2^{-n/\mu'}\).\]
	\end{theorem}
	\begin{IEEEproof}
		Section~\ref{sec:sketch} sketchs the proof.
		Section~\ref{sec:proof} details the proof.
		See Appendix~\ref{app:plane} for comparison.
	\end{IEEEproof}
	
	\begin{theorem} \label{thm:true}
		Assume the Scaling Assumption \cite[Formula~(12)]{FV14},
		and the consequence that $\mu=3.627$.
		Then there exists $h$ in the sense of Theorem~\ref{thm:hmu}
		such that $\mu^*$ is arbitrarily close to $\mu=3.627$.
	\end{theorem}
	\begin{IEEEproof}
		See Appendix~\ref{pf:true}.
	\end{IEEEproof}
	
	\begin{corollary} \label{cor:recover}
		Theorem~\ref{thm:main} recovers the scaling exponent as a special case.
	\end{corollary}
	\begin{IEEEproof}
		See Appendix~\ref{pf:recover}.
	\end{IEEEproof}
	
	\begin{corollary} \label{cor:outperform}
		Theorem~\ref{thm:main} implies Theorem~\ref{thm:gamma}
		(i.e.,~\cite[Theorem~7]{MHU16}) as a special case.
		It recovers the error exponent as a special case.
	\end{corollary}
	\begin{IEEEproof}
		See Appendix~\ref{pf:outperform}.
	\end{IEEEproof}
%	\fermat{Rethink how to rearrange these corollary.}

\section{Sketch of Proof of Theorem \ref{thm:main}} \label{sec:sketch}

	The complete proof is in Section~\ref{sec:proof}.
	
	We will attempt to move upward $n_0\coloneqq\<n\mu^*/\mu'\>$ steps or less
%	\fermat{$\<\>$ is defined later.}
	and to move rightward $n_1\coloneqq n-n_0$ steps or more.

\subsection{Discretization: Calculate the Number of Pockets}

	We will be using $D$ pockets
%	\fermat{What is the correct way to line-break a math list?}
	\[\AA n{n_0/D},\AA n{2n_0/D},\AA n{3n_0/D},\dotsc,\AA n{Dn_0/D}.\]
	The tighter the Formula~(\ref{eq:prere}) is the more pockets we need.
	Let $m$ be between $0$ and $n_0$ that indexes the pockets.

\subsection{Multi-Pocket Recruit Phase}

	Pocket $\A nm$ collects synthetic channels $\W{2^m}j$
	with erasure probability less than $P\ub2^{-Dm}$
	whose ancestors are not collected by any pocket with smaller index.
	We give each synthetic channel $\W{2^m}j$ a weight of $2^{-m}$.
	Then pocket $\A nm$ will weigh $2^{-m/\mu^*+n_0/D\mu^*}$.
%	\fermat{no pounds, verb is weigh}

\subsection{Multi-Pocket Train Phase}

	For each synthetic channel $\W{2^m}j$ in pocket $\A nm$,
	replace it with all its descendants of the form $\W{2^n}{2^{n-m}j-k}$.

\subsection{Multi-Pocket Retain Phase}
	
	Claim a threshold $\epsilon=\beta'n/(n-m)$.
	For each synthetic channel $\W{2^n}{2^{n-m}j-k}$
	obtained from $\W{2^m}j$ in pocket $\A nm$,
	discard it if its erasure probability is more than $2^{2^{-\beta'n}}$.

\subsection{Estimate the Error Probability}

	By how we discard synthetic channels in the retain phase,
	the block error probability will be less than $2^n\cdot2^{2^{\beta'n}}$.

\subsection{Estimate the Gap to Capacity}

	Pocket $\A nm$ weighs $2^{-m/\mu^*+n_0/D\mu^*}$ in the recruit phase
	and loses $2^{-(n-m)(1-H_2(\epsilon))}$ of its weight in the retain phase.
	So it loses only $2^{-m/\mu^*+n_0/D\mu^*}\cdot2^{-(n-m)(1-H_2(\epsilon))}$
	units of weight.
	This quantity is $O\(2^{-n/\mu'}\)$ by Formula~(\ref{eq:prere}).

\subsection{Summary}

	Hence the union 
	\[\clA_n\coloneqq\AA n{n_0/D}\cup\AA n{2n_0/D}\cup\dotsb\cup\A n{Dn_0/D}\]
	will have gap to capacity $O\(2^{-n/\mu'}\)$
	and block error probability less than $2^n\cdot2^{2^{-\beta'n}}$.
	
	The complete proof is in the next section.

\section{Proof of Theorem \ref{thm:main}} \label{sec:proof}

\subsection{Discretization: Calculate the Number of Pockets}

	By continuity, there exists a positive integer $D>0$ such that
	\[\frac{1-(\pi+\delta_1)}{\mu'-\mu^*(\pi+\delta_2)}+
		H_2\(\frac{\beta'\mu'}{\mu'-\mu^*(\pi+\delta_3)}\)
		<1 \label{eq:perturb}\]
	for all $-9/D<\delta_1,\delta_2,\delta_3<9/D$.
	We are going to use about $D$ pockets and apply Lemma~\ref{lem:D} with
	$D=D$.
	
	In the following proof, expressions like $\<2n_0/D\>$ and $\<n\mu^*/\mu'\>$
	are meant to be integers that are very close to the real numbers
	$2n_0/D$ and $n\mu^*/\mu'$.
	It does not matter whether we round up or round down, as we will see later
	that Formula~(\ref{eq:perturb}) permits such flexibility.

\subsection{Multi-Pocket Recruit Phase} \label{sec:recruit}

	Let $P\ub$ be a small, but fixed number.
	Let $n$ be large enough.
	Let $n_0\coloneqq\<n\mu^*/\mu'\>$.
%	\fermat{$n_1$ is not used}
	We define the pockets
	\[\AA n{n_0/D},\AA n{2n_0/D},\AA n{3n_0/D},\dotsc,\AA n{Dn_0/D}\]
	by letting $\A nm$ collect synthetic channels $\W{2^m}j$
	with erasure probability less than $P\ub2^{-Dm}\le P\ub2^{-n_0}$.
	
	We give each synthetic channel $\W{2^m}j$ a weight of $2^{-m}$.
%	\fermat{Oops, imperial units.}
	Pocket $\A nm$ should weigh $I(W)-O\(2^{-m/\mu^*}\)$
	by Lemma~\ref{lem:D} with $D=D$.
	
	For each pocket $\A nm$, discard synthetic channels that has some ancestor
	collected in a pocket with smaller index
	because we do not want to double-count.
	Now the union $\AA n{n_0/D}\cup\AA n{2n_0/D}\cup\dotsb\cup\A nm$ weighs
	between $I(W)-O\(2^{-m/\mu^*}\)$ and $I(W)+O(2^{-n_0})$.
	(The upper bound comes from entropy conservation.)
	Thus $\A nm$ weighs at most $O\(2^{-m/\mu^*+n_0/D\mu^*}\)$.

\subsection{Multi-Pocket Train Phase} \label{sec:train}

	For each synthetic channel $\W{2^m}j$ in pocket $\A nm$,
	replace it with all its descendants of the form $\W{2^n}{2^{n-m}j-k}$.
	Doing this does not affect the weight of $\A nm$.
	
	For each descendant $\W{2^n}{2^{n-m}j-k}$
	obtained from $\W{2^m}j$ in pocket $\A nm$,
	its erasure probability is doubled and squared totally $n-m$ times.

\subsection{Multi-Pocket Retain Phase} \label{sec:retain}

	Claim a threshold $\epsilon=\beta'n/(n-m)$.
	For each synthetic channel $\W{2^n}{2^{n-m}j-k}$
	obtained from $\W{2^m}j$ in pocket $\A nm$,
	discard it if its erasure probability is squared less than
	$\beta'n=\epsilon(n-m)$ times out of $n-m$ chances.
%	\fermat{Note that $\beta'n$ is not necessary an integer,
%		but the sentence still makes sense.}
	By \cite[Formula~(1.59)]{RU08} with $\epsilon=\epsilon$,
	pocket $\A nm$ loses at most $2^{-(n-m)(1-H_2(\epsilon))}$ of its weight
	here.
	
	Furthermore, discard synthetic channels
	with erasure probability more than $2^{2^{-\beta'n}}$.
	By \cite[Lemma~22]{HAU14} with $x\coloneqq P\ub2^{-n_0}$,
	pocket $\A nm$ loses
	\[O\(x(1-\log x)\)=O\(P\ub2^{-n_0}(1+n_0)\)\]
	of its weight.
	If $n,n_0$ are large enough, the weight loss is $2^{-n_0(1-o(1))}$.
	This quantity is much much smaller than the targeted gap to capacity
	$O\(2^{-n/\mu'}\)=O\(2^{-n_0/\mu^*}\)$ so we will simply ignore this.

\subsection{Estimate the Error Probability}

	By how we discard synthetic channels in the retain phase
	(Section~\ref{sec:retain}), the synthetic channels in the union
	$\AA n{n_0/D}\cup\AA n{2n_0/D}\cup\dotsb\cup\AA n{Dn_0/D}$
	have their erasure probabilities less than $2^{2^{\beta'n}}$.
	By union bound, the block error probability is less than
	$2^n\cdot2^{2^{\beta'n}}$.

\subsection{Estimate the Gap to Capacity}

	Now we try to weigh the union
	$\AA n{n_0/D}\cup\AA n{2n_0/D}\cup\dotsb\cup\A n{Dn_0/D}$.
	
	In the recruit phase (Section~\ref{sec:recruit}), the union
	weighs at least $I(W)-O(2^{n_0/\mu^*})=I(W)-O(2^{-n/\mu'})$;
	and each $\A nm$ weighs at most $O\(2^{-m/\mu^*+n_0/D\mu^*}\)$.
	
	And then in the train phase (Section~\ref{sec:train}) the weight remains.
	
	Finally in the retain phase (Section~\ref{sec:retain}), each $\A nm$
	loses at most $2^{-(n-m)(1-H_2(\epsilon))}$ of its weight.
	Thus it loses at most
	\[O\(2^{-m/\mu^*+n_0/D\mu^*}\)\cdot2^{-(n-m)(1-H_2(\epsilon))}\]
	units of weight.
%	\fermat{Oh, no... pounds?}
	Take logarithm.
	Each $\A nm$ loses $2$ to the power of
%	\fermat{Why $2$ here?}
	\[-\frac m{\mu^*}+\frac{n_0}{D\mu^*}
		-(n-m)\(1-H_2\(\frac{\beta'n}{n-m}\)\)+O(1)\]
	units of weight.
	
	Recall that $0<m\le n_0$.
	Let $\pi$ be such that $m=n\pi\mu^*/\mu'$,
	so $m=n_0\pi+O(1)$ and $n-m=n(\mu'-\pi\mu^*)/\mu'$.
%	\fermat{Be careful, rounding error.}
	If $n,n_0$ are large enough then $\pi\in[-1/D,1+1/D]$.
	Now the main term of the logarithm becomes
	\[-\frac m{\mu^*}+\frac{n_0}{D\mu^*}-(n-m)
		\(1-H_2\(\frac{\beta'\mu'}{\mu'-\pi\mu^*}\)\).\]
	By Formula~(\ref{eq:perturb}) this quantity is less than
	\[-\frac m{\mu^*}+\frac{n_0}{D\mu^*}
		-(n-m)\frac{1-(\pi-1/D)}{\mu'-\mu^*\pi}\]
	which, up to constants, is equal to
	\[-\frac{n(\pi-1/D)}{\mu'}-\frac{n-n(\pi-1/D)}{\mu'}=\frac{-n}{\mu'}.\]
	Hence each $\A nm$ loses at most $O\(2^{-n/\mu'}\)$ units of weight.
	Hence the union loses at most $DO\(2^{-n/\mu'}\)=O\(2^{-n/\mu'}\)$
	units of weight, which still weighs
	\[I(W)-O\(2^{-n/\mu'}\).\]

\subsection{Summary}

	To summarize, the union 
	\[\clA_n\coloneqq\AA n{n_0/D}\cup\AA n{2n_0/D}\cup\dotsb\cup\A n{Dn_0/D}\]
	has gap to capacity
	\[O\(2^{-n/\mu'}\)\]
	and block error probability less than
	\[2^n\cdot2^{2^{-\beta'n}}.\]
	This completes the proof of Theorem~\ref{thm:main}.

\section{Future Works}

\subsection{Regarding Eigenfunction}

	Lemma~\ref{lem:alpha} plays the same role in proving Theorem~\ref{thm:main}
	as that \cite[Lemma~2]{AT09} plays in proving \cite[Theorem~3]{AT09}
	and that \cite[Lemma~5]{MHU16} plays in proving \cite[Theorem~7]{MHU16}.
	The three lemmata provide some ``initial boost''
	before applying the ``doubling-or-squaring'' argument
	(i.e.,~the train phase and retain phase).
%	\fermat{Fixit}
	
	They are, in contrast to the ``doubling-squaring'' argument,
	a pretty weak starting point.
	But from the main theorem
	(Theorem~\ref{thm:main} and Appendix~\ref{app:plane})
	we know that the initial boost could be doubly exponential in $n$.
	A potential proof will be to consult function $h$ in Theorem~\ref{thm:hmu}
	its behavior near $\xi=0$.
	
	In particular: Is there $h,\mu$ such that
	\[\frac{h\(\xi^2\)+h\(2\xi-\xi^2\)}{2h(\xi)}=2^{-1/\mu}?\]
	(Equivalently \cite[Formula~(12)]{FV14}.)
	If so, could
%	\fermat{Pick a good greek letter.}
	\[h(\xi)\propto\(-\log\xi\)^{-\theta}\text{ for }
		\theta\coloneqq-\log_2\(2^{1-1/\mu}-1\)?\]
	One may notice
%	\fermat{Probably not... hahaha}
	that when $\mu=3.627$, the number $1/\mu\theta\approx.4469$
	is in Appendix~\ref{app:plane}.
	That is to say, such infinitesimal behavior of $h$
	implies the straight segment from $(0,1/\mu)$ to $(1/\mu\theta,0)$.

\subsection{Regarding Convex Hull}

	The next question is whether moving upward and moving rightward
	follow vector addition.
	If so, then it trivially implies the straight segment from
	$(0,1/\mu)$ to $(\beta,0)$.
	Moreover, do there exist achievable points beyond that segment?
%	\fermat{Add your question here}

\subsection{Regarding General Channels}

	We have not said anything about binary symmetric memoryless channels
	but we are confident that there are similar results.
	The reasons are that the scaling exponent is well-defined for other channels
	and that the ``doubling-squaring'' phenomenon is simply omnipresent.

\section{Concluding Remarks}

	We investigate the trading-off between block length, code rate,
	and block error probability in constructing classical polar codes.
	
	Our result, Theorem~\ref{thm:main}, specializes
	to the result that the error exponent is $\beta=1/2$
	\cite[Theorem~1]{AT09} by Corollary~\ref{cor:outperform}.
	and to the result that the scaling exponent is $\mu=3.627$
	\cite[Abstract]{FV14} by Corollary~\ref{cor:recover}.
	
	Moreover, our result implies all known trading-off results:
	mainly \cite[Theorem~1]{GX13} and \cite[Theorem~7]{MHU16}
	by Corollary~\ref{cor:outperform}.
%	\fermat{How to typeset this?
%		``Especially''?}
%	\fermat{Note that \cite[Theorem~7]{MHU16} = Theorem~\ref{thm:gamma}.}
	
	It remains open whether there is room for improvement or not.
%	\fermat{Correct grammar?}

\bibliographystyle{IEEEtran}
\bibliography{Exponent-4}

\begin{thebibliography}{10}
\providecommand{\url}[1]{#1}
\csname url@samestyle\endcsname
\providecommand{\newblock}{\relax}
\providecommand{\bibinfo}[2]{#2}
\providecommand{\BIBentrySTDinterwordspacing}{\spaceskip=0pt\relax}
\providecommand{\BIBentryALTinterwordstretchfactor}{4}
\providecommand{\BIBentryALTinterwordspacing}{\spaceskip=\fontdimen2\font plus
\BIBentryALTinterwordstretchfactor\fontdimen3\font minus
  \fontdimen4\font\relax}
\providecommand{\BIBforeignlanguage}[2]{{%
\expandafter\ifx\csname l@#1\endcsname\relax
\typeout{** WARNING: IEEEtran.bst: No hyphenation pattern has been}%
\typeout{** loaded for the language `#1'. Using the pattern for}%
\typeout{** the default language instead.}%
\else
\language=\csname l@#1\endcsname
\fi
#2}}
\providecommand{\BIBdecl}{\relax}
\BIBdecl

\bibitem{Arikan09}
E.~Arikan, ``Channel polarization: A method for constructing capacity-achieving
  codes for symmetric binary-input memoryless channels,'' \emph{IEEE
  Transactions on Information Theory}, vol.~55, no.~7, pp. 3051--3073, July
  2009.

\bibitem{MHU16}
M.~Mondelli, S.~H. Hassani, and R.~L. Urbanke, ``Unified scaling of polar
  codes: Error exponent, scaling exponent, moderate deviations, and error
  floors,'' \emph{IEEE Transactions on Information Theory}, vol.~62, no.~12,
  pp. 6698--6712, Dec 2016.

\bibitem{AT09}
E.~Arikan and E.~Telatar, ``On the rate of channel polarization,'' in
  \emph{2009 IEEE International Symposium on Information Theory}, June 2009,
  pp. 1493--1495.

\bibitem{Gallager65}
R.~Gallager, ``A simple derivation of the coding theorem and some
  applications,'' \emph{IEEE Transactions on Information Theory}, vol.~11,
  no.~1, pp. 3--18, January 1965.

\bibitem{Gallager68}
R.~G. Gallager, \emph{Information Theory and Reliable Communication}.\hskip 1em
  plus 0.5em minus 0.4em\relax New York, NY, USA: John Wiley \& Sons, Inc.,
  1968.

\bibitem{KSU10}
S.~B. Korada, E.~Sasoglu, and R.~Urbanke, ``Polar codes: Characterization of
  exponent, bounds, and constructions,'' \emph{IEEE Transactions on Information
  Theory}, vol.~56, no.~12, pp. 6253--6264, Dec 2010.

\bibitem{HMTU13}
S.~H. Hassani, R.~Mori, T.~Tanaka, and R.~L. Urbanke, ``Rate-dependent analysis
  of the asymptotic behavior of channel polarization,'' \emph{IEEE Transactions
  on Information Theory}, vol.~59, no.~4, pp. 2267--2276, April 2013.

\bibitem{MT14}
R.~Mori and T.~Tanaka, ``Source and channel polarization over finite fields and
  reed-solomon matrices,'' \emph{IEEE Transactions on Information Theory},
  vol.~60, no.~5, pp. 2720--2736, May 2014.

\bibitem{KMTU10}
S.~B. Korada, A.~Montanari, E.~Telatar, and R.~Urbanke, ``An empirical scaling
  law for polar codes,'' in \emph{2010 IEEE International Symposium on
  Information Theory}, June 2010, pp. 884--888.

\bibitem{HAU14}
S.~H. Hassani, K.~Alishahi, and R.~L. Urbanke, ``Finite-length scaling for
  polar codes,'' \emph{IEEE Transactions on Information Theory}, vol.~60,
  no.~10, pp. 5875--5898, Oct 2014.

\bibitem{GB14}
D.~Goldin and D.~Burshtein, ``Improved bounds on the finite length scaling of
  polar codes,'' \emph{IEEE Transactions on Information Theory}, vol.~60,
  no.~11, pp. 6966--6978, Nov 2014.

\bibitem{FV14}
A.~Fazeli and A.~Vardy, ``On the scaling exponent of binary polarization
  kernels,'' in \emph{2014 52nd Annual Allerton Conference on Communication,
  Control, and Computing (Allerton)}, Sept 2014, pp. 797--804.

\bibitem{PU16}
H.~D. Pfister and R.~Urbanke, ``Near-optimal finite-length scaling for polar
  codes over large alphabets,'' in \emph{2016 IEEE International Symposium on
  Information Theory (ISIT)}, July 2016, pp. 215--219.

\bibitem{Dobrushin61}
\BIBentryALTinterwordspacing
R.~L. Dobrushin, ``Mathematical problems in the shannon theory of optimal
  coding of information,'' in \emph{Proceedings of the Fourth Berkeley
  Symposium on Mathematical Statistics and Probability, Volume 1: Contributions
  to the Theory of Statistics}.\hskip 1em plus 0.5em minus 0.4em\relax
  Berkeley, Calif.: University of California Press, 1961, pp. 211--252.
  [Online]. Available: \url{https://projecteuclid.org/euclid.bsmsp/1200512168}
\BIBentrySTDinterwordspacing

\bibitem{Strassen64}
\BIBentryALTinterwordspacing
V.~Strassen, ``Asymptotische absch{\"a}tzungen in shannons
  informationstheorie,'' in \emph{Transactions of the Third Prague Conference
  on Information Theory}.\hskip 1em plus 0.5em minus 0.4em\relax Publishing
  House of the Czechoslovak Academy of Sciences, 1962, pp. 689--723. [Online].
  Available: \url{https://www.math.cornell.edu/~pmlut/strassen.pdf}
\BIBentrySTDinterwordspacing

\bibitem{TZ00}
\BIBentryALTinterwordspacing
J.-P. Tillich and G.~Z\'emor, ``Discrete isoperimetric inequalities and the
  probability of a decoding error,'' \emph{Combin. Probab. Comput.}, vol.~9,
  no.~5, pp. 465--479, 2000. [Online]. Available:
  \url{https://doi.org/10.1017/S0963548300004466}
\BIBentrySTDinterwordspacing

\bibitem{Montanari01}
\BIBentryALTinterwordspacing
A.~Montanari, ``Finite-size scaling and metastable states of good codes,'' in
  \emph{Proceedings of the Allerton Conference on Communication, Control and
  Computing}, Oct 2001. [Online]. Available:
  \url{https://web.stanford.edu/~montanar/RESEARCH/FILEPAP/allerton01.pdf}
\BIBentrySTDinterwordspacing

\bibitem{Hayashi09}
M.~Hayashi, ``Information spectrum approach to second-order coding rate in
  channel coding,'' \emph{IEEE Transactions on Information Theory}, vol.~55,
  no.~11, pp. 4947--4966, Nov 2009.

\bibitem{PPV10}
Y.~Polyanskiy, H.~V. Poor, and S.~Verdu, ``Channel coding rate in the finite
  blocklength regime,'' \emph{IEEE Transactions on Information Theory},
  vol.~56, no.~5, pp. 2307--2359, May 2010.

\bibitem{Hassani13}
\BIBentryALTinterwordspacing
S.~H. Hassani, ``Polarization and spatial coupling: Two techniques to boost
  performance,'' \emph{Ecole Polytechnique Federale de Lausanne}, no. 5706,
  2013. [Online]. Available: \url{https://infoscience.epfl.ch/record/188275/}
\BIBentrySTDinterwordspacing

\bibitem{FHMV17}
\BIBentryALTinterwordspacing
A.~Fazeli, S.~H. Hassani, M.~Mondelli, and A.~Vardy, ``Binary linear codes with
  optimal scaling and quasi-linear complexity,'' \emph{CoRR}, vol.
  abs/1711.01339, 2017. [Online]. Available:
  \url{http://arxiv.org/abs/1711.01339}
\BIBentrySTDinterwordspacing

\bibitem{MHU15}
M.~Mondelli, S.~H. Hassani, and R.~L. Urbanke, ``Scaling exponent of list
  decoders with applications to polar codes,'' \emph{IEEE Transactions on
  Information Theory}, vol.~61, no.~9, pp. 4838--4851, Sept 2015.

\bibitem{GX13}
V.~Guruswami and P.~Xia, ``Polar codes: Speed of polarization and polynomial
  gap to capacity,'' in \emph{2013 IEEE 54th Annual Symposium on Foundations of
  Computer Science}, Oct 2013, pp. 310--319.

\bibitem{BGNRS18}
\BIBentryALTinterwordspacing
J.~Blasiok, V.~Guruswami, P.~Nakkiran, A.~Rudra, and M.~Sudan, ``General strong
  polarization,'' \emph{CoRR}, vol. abs/1802.02718, 2018. [Online]. Available:
  \url{http://arxiv.org/abs/1802.02718}
\BIBentrySTDinterwordspacing

\bibitem{BPT13}
M.~B. Parizi and E.~Telatar, ``On the correlation between polarized becs,'' in
  \emph{2013 IEEE International Symposium on Information Theory}, July 2013,
  pp. 784--788.

\bibitem{RU08}
\BIBentryALTinterwordspacing
T.~Richardson and R.~Urbanke, \emph{Modern coding theory}.\hskip 1em plus 0.5em
  minus 0.4em\relax Cambridge University Press, Cambridge, 2008. [Online].
  Available: \url{https://doi.org/10.1017/CBO9780511791338}
\BIBentrySTDinterwordspacing

\bibitem{Durrett10}
\BIBentryALTinterwordspacing
R.~Durrett, \emph{Probability: Theory and Examples}, 4th~ed.\hskip 1em plus
  0.5em minus 0.4em\relax New York, NY, USA: Cambridge University Press, 2010.
  [Online]. Available: \url{https://services.math.duke.edu/~rtd/PTE/PTE4_1.pdf}
\BIBentrySTDinterwordspacing

\end{thebibliography}

\appendix

\subsection{Proof of Lemma~\ref{lem:alpha}} \label{pf:alpha}

	Fix an $m$, there are three events that partition the sample space.
	\begin{align}
		S &\coloneqq \Set{Z_m\le P\ub2^{-Dm}}; \\
		M &\coloneqq \Set{P\ub2^{-Dm}<Z_m<1-P\ub2^{-Dm}}; \\
		L &\coloneqq \Set{1-P\ub2^{-Dm}\le Z_m}.
	\end{align}
	Here $M$ is equivalent to
	\[\Set{Z_m(1-Z_m)>P\ub2^{-Dm}(1-P\ub2^{-Dm})}\]
	and to
	\[*\Set{\(Z_m(1-Z_m)\)^\alpha>\(P\ub2^{-Dm}(1-P\ub2^{-Dm})\)^\alpha}.\]*
	By Markov's inequality \cite[Theorem~1.6.4]{Durrett10}
	\[\bbP(M)\le\frac{\bbE\left[\(Z_m(1-Z_m)\)^\alpha\right]}
		{\(P\ub2^{-Dm}(1-P\ub2^{-Dm})\)^\alpha}.\]
	Thus, provided that $P\ub$ is small enough,
	\[\bbP(M)\le\frac{c_12^{-m\rho}}{2^{-Dm\alpha}}=O\(2^{-m(\rho+D\alpha)}\).\]
	Also by Markov's inequality
	\begin{align}
		\bbP(L) &\ge \frac{\bbE[Z_m]}{1-P\ub2^{-Dm}}=\bbE[Z_m]+O\(2^{-Dm}\) \\
		&= Z_0+O\(2^{-Dm}\)=1-I(W)+O\(2^{-Dm}\)
	\end{align}
	Now $\bbP\(Z_m\le P\ub2^{-Dm}\)$
	\begin{align}
		&= \bbP(S) \\
		&= 1-\bbP(L)-\bbP(M) \\
		&\le 1-Z_0-O\(2^{-Dm}\)-O\(2^{-m(\rho+D\alpha)}\) \\
		&= I(W)-O\(2^{-m(\rho+D\alpha)}\).
	\end{align}

\subsection{Proof of Lemma~\ref{lem:D}} \label{pf:D}

	Note that Formula~(\ref{eq:hmu}) is an inequality,
	so there exists $\rho_1$ larger than $1/\mu^*$
	such that Formula~(\ref{eq:hrho1}) in Lemma~\ref{lem:hrho1} holds.
	So Formula~(\ref{eq:delta}),
	the conclusion of Lemma~\ref{lem:hrho1}, holds.
	
	Choose a $\rho$ that lies between $\rho_1$ and $1/\mu^*$.
	Choose a small $\alpha$ such that $\rho-D\alpha>1/\mu^*$.
	Choose a small $\delta$ such that the right hand side of
	Formula~(\ref{eq:delta}) is $O(2^{-m\rho})$, i.e, make
	\[\bbE\left[\(Z_m(1-Z_m)\)^\alpha\right]\le O(2^{-m\rho})\]
	hold.
	
	Now apply Lemma~\ref{lem:alpha} with $(\alpha,\rho)=(\alpha,\rho)$.
	Its conclusion implies
	\[\bbP\(Z_m\le P\ub2^{-Dm}\)\ge I(W)-O\(2^{-m/\mu^*}\).\]
	This completes the proof.

\subsection{Proof of Theorem~\ref{thm:true}} \label{pf:true}

	The assumption says that there exists $f(z,a,b)$ such that
	\[f(\xi,a,b)=\lim_{n\to\infty}2^{n/\mu}f_n(\xi,a,b)\]
	where
	\[f_{n+1}(\xi,a,b)=\frac{f_n\(\xi^2,a,b\)+f_n\(2\xi-\xi^2,a,b\)}2.\]
	
	To prove the theorem,
	take whatever $a<b$ and let $h(\xi)\coloneqq f(\xi,a,b)$.
	Then $2^{-1/\mu}h(\xi)$
	\begin{align}
		&= 2^{-1/\mu}\lim_{n+1\to\infty}2^{(n+1)/\mu}f_{n+1}(\xi,a,b) \\
		&= \lim_{n\to\infty}2^{n/\mu}
			\frac{f_n\(\xi^2,a,b\)+f_n\(2\xi-\xi^2,a,b\)}2 \\
		&= \frac12\biggl(\lim_{n\to\infty}2^{n/\mu}f_n\(\xi^2,a,b\) \\
		&\mkern80mu
			+\lim_{n\to\infty}2^{n/\mu}f_n\(2\xi-\xi^2,a,b\)\biggr) \\
		&= \frac{h\(\xi^2\)+h\(2\xi-\xi^2\)}2.
	\end{align}
	That is, Formula~(\ref{eq:hmu}) is equality for $\mu^*=\mu$.
	Thus the inequality holds for $\mu^*$ arbitrarily close to $\mu$.

\subsection{Proof of Corollary~\ref{cor:recover}} \label{pf:recover}

	Apply Theorem~\ref{thm:main} with
%	\fermat{Elaborate this.}
	\[\mu'\coloneqq\frac{\mu^*}{1-\gamma},\qquad\beta'\coloneqq\beta_*\gamma\]
	for some fixed $\beta_*\le.4469$ and $\mu^*>\mu$ and all $\gamma\in[0,1]$.
	In detail: Choose, for instance, $\beta_*\coloneqq.4469$.
	It is easy (numerical) to verify that
	\[\frac{1-\xi}\mu+H_2\(\beta_*\xi\)<1\]
	for $\xi\in[0,1]$ and $\mu=3.627$.
	Consequently
	\[\frac{1-\xi}{\mu^*}+H_2\(\beta_*\xi\)<1\]
	for $\mu^*>\mu$.
	With $\xi\coloneqq\gamma/(1-\pi+\pi\gamma)$ this becomes
	\[\frac{1-\pi-\gamma-\pi\gamma}{\mu^*(1-\pi+\pi\gamma)}
		+H_2\(\frac{\beta_*\gamma}{1-\pi+\pi\gamma}\)<1\]
	This is exactly Formula~(\ref{eq:prere})
	\[*\frac{1-\pi}{\mu'-\mu^*\pi}+H_2\(\frac{\beta'\mu'}{\mu'-\mu^*\pi}\)<1\]*
	with the corresponding $\mu'$ and $\beta'$.
	
	Now apply Theorem~\ref{thm:true} with
	\[\gamma\to0,\qquad\mu^*\to\mu.\]

\subsection{Proof of Corollary~\ref{cor:outperform}} \label{pf:outperform}

	To recover Theorem~\ref{thm:gamma}, plug Formula~(\ref{eq:gamma})
	\[*\beta'\coloneqq\gamma H_2^{-1}\(\frac{\gamma(\mu^*+1)-1}{\gamma\mu^*}\)
		\text{ and }\mu'\coloneqq\frac{\mu^*}{1-\gamma}\]*
	in Formula~(\ref{eq:prere})
	\[*\frac{1-\pi}{\mu'-\mu^*\pi}+H_2\(\frac{\beta'\mu'}{\mu'-\mu^*\pi}\)<1\]*
	and verify it.

\newpage

	In detail:
	First with $0\le\pi\le1$, the first term of the inequality is
	\[\frac{1-\pi}{\mu'-\mu^*\pi}\le\frac1{\mu'}
		=\frac{1-\gamma}{\mu^*}=\frac{\gamma-\gamma^2}{\gamma\mu^*}.\]
	Again with $0\le\pi\le1$ and $H_2,H_2^{-1}$ monotonically increasing,
	the second term of the inequality is
	\begin{align}
		H_2\(\frac{\beta'\mu'}{\mu'-\mu^*\pi}\)
		&\le H_2\(\frac{\beta'\mu'}{\mu'-\mu^*}\)=H_2\(\frac{\beta'}\gamma\) \\
		&= H_2\(H_2^{-1}\(\frac{\gamma(\mu^*+1)-1}{\gamma\mu^*}\)\) \\
		&=\frac{\gamma(\mu^*+1)-1}{\gamma\mu^*}.
	\end{align}
	So the left hand side of the inequality is
	\[\frac{\gamma-\gamma^2}{\gamma\mu^*}+\frac{\gamma(\mu^*+1)-1}{\gamma\mu^*}
		=\frac{\gamma\mu^*-(1-\gamma)^2}{\gamma\mu^*}<1.\]
	
	P.S. We know it works because multi-pocket trick implies one-pocket trick.
	And the one-pocket trick is a generalization of \cite[Proposition~3]{AT09}.
	(See \cite[Formula~(31)]{AT09}.)
	
	Now Theorem~\ref{thm:gamma} recovers the error exponent as a special case
	by driving $\gamma\to1$.

\vfill
\centerline{The rest of this page is intensionally left black.}
\vfill

%\AtBeginShipout{\AtBeginShipoutUpperLeft{
%	\tikz[rp,o]\draw[yellow](current page text area.south west)
%		           rectangle(current page text area.north east);}}

\clearpage
\mbox{}
\newpage

\subsection{Visualization of \texorpdfstring{$\mu$}{μ}} \label{app:muruler}

{\makeatletter
	Unless otherwise stated, we assume binary alphabet,
	classical kernel, and binary erasure channel.
	See Section~\ref{sec:ser} for details.
	\def\readpair#1,#2,{\xdef\xcoord{#1}\xdef\ycoord{#2}}
	\def\ymax{1/2}
	\def\ymin{1/6}
	\pgfmathsetmacro\yscale{(\textheight-\baselineskip)/(\ymax-\ymin)}
	\tikzdeclarecoordinatesystem{mu}{
		\readpair#1,
		\tikz@scan@one@point\relax(current page text area.south west)
		\pgfpoint{\pgf@x+1.2cm+\xcoord cm}
			{\pgf@y+.5ex+(1/\ycoord-\ymin)*\yscale}
	}
	\tikz[rp,o]\draw(mu cs:0,2)--(mu cs:0,6)
		foreach\y in{2,2.25,...,6}{(mu cs:-.1,\y)node[left]{$1/\y$}--+(.2,0)};
%	\fermat{Fix spacing, TikZ is currently here.}
%	\fermat{line color???}
	
	\leftskip1cm
	\advance\rightskip0ptplus1fil
	\parindent-1cm
	\def\larger{\textcolor{red}{\LARGE larger}\ }
	\catcode`\@13
	\def@#1#2 {\vfill\par\tikz[rp,o,b=-.5ex]\draw%[blue]
			(0,0)--(mu cs:.1,#2)--+(-.2,0); $#1#2$: }
	@\to2 \cite{PU16} \larger kernels over \larger alphabets
		achieve optimal exponent
	@\to2 \cite{Hassani13} conjectures that \larger kernels
		over binary alphabet suffice.
	@\to2 \cite{FHMV17} \larger (random) kernels suffice
	@=3.356 \cite{FV14} a \larger kernel of size $16$
	@=3.577 \cite{FV14} a \larger kernel of size $8$
	@\ge3.579 \cite{HAU14} for general channels
	@\approx3.6261 \cite{KMTU10} empirically
	@\approx3.627 \cite{HAU14} conjectures this value
	@=3.627 \cite{FV14}
	@\le3.639 \cite{MHU16}
	@\le4.714 \cite{MHU16} for general channels
	@\le5.702 \cite{GB14} for general channels
	@\le6     \cite{HAU14} for general channels
	
}

\clearpage

\mbox{}
\newpage

\subsection{Visualization of \texorpdfstring{$\beta$}{β}} \label{app:betaruler}

{\makeatletter
	See Section~\ref{sec:eer} for details.
	
	\def\readpair#1,#2,{\xdef\xcoord{#1}\xdef\ycoord{#2}}
	\def\mlef{1.7}\def\xmin{.5}\def\xmax{1}\def\mrig{1.7}
	\pgfmathsetmacro\xscale{(21-\mlef-\mrig)/(\xmax-\xmin)}
	\tikzdeclarecoordinatesystem{beta}{
		\readpair#1,
		\tikz@scan@one@point\relax(current page.center)
		\pgf@xa\pgf@x
		\tikz@scan@one@point\relax(pic cs:yref)
		\pgfmathsetlength\pgf@x
			{\pgf@xa-10.5cm+\mlef cm+(\xcoord-\xmin)*\xscale cm}
		\pgfmathsetlength\pgf@y{\pgf@y+0cm+\ycoord cm}
	}
	\tikz[rp,o,b=-.5ex]\draw(beta cs:.5,0)--(beta cs:1,0)
		foreach\x in{10,...,20}{(beta cs:\x/20,-.1)node[below]
			{$\expandafter\Pgf@geT\the\dimexpr\x pt/20$}--+(0,.2)};
	\advance\rightskip0ptplus1fil
	\parindent0pt
	\catcode`\@13
	\def@#1#2 {\par\tikz[rp,o]\draw%[blue]
			(0,0)-|(beta cs:#2,.1)--+(0,-.2); $#1#2$: }
	@=.5 \cite{AT09}
	@=.5 \cite{HMTU13} second order term given
	@=.51828 \cite{KSU10} a $16$-by-$16$ kernel
	@=.52205 \cite{KSU10} a $30$-by-$30$ kernel
	@=.52643 \cite{KSU10} a $31$-by-$31$ kernel
	\def@#1#2 #3@{\par
		\mbox{} $#1#2$: #3
		\tikz[rp,o]\draw%[blue]
			(0,0)-|(beta cs:#2,.1)--+(0,-.2);}
	@\to1 \cite{KSU10} larger kernel @
	@\to1 \cite{MT14} larger alphabet and larger kernel @
	
	\tikzmark{yref}
}
\vskip5.5cm

\subsection{Visualization of Moderate Deviation} \label{app:plane}

	The following plot assumes the Scaling Assumption \cite[Formula~(12)]{FV14}
	and $\mu=3.627=3627/1000$ \cite[Abstract]{FV14}.
	\fermat{It is now more informative but more messy.
		Be careful.}
	See Section~\ref{sec:mdr} for details.
	\fermat{Space from bottom?}
	\fermat{Coloring?}
	
	\everymath{\displaystyle}
	\tikzset{
		bad/.style={brown},
		Our/.style={teal},
		MHU16/.style={violet},
	}
	\tikz[rp,o,shift=(current page text area.south),shift={(-6.7,1)},scale=3]{
		\draw[->](0,0)--(0,2.9)node[above right,xshift=-68]
			{$y\text{-coordinate}=\liminf_{n\to\infty}\frac
			{-\log(\text{Gap to capacity})}{\log(\text{Block Length})}$};
		\draw[->](0,0)--(5.2,0)node[below left,xshift=-90,black]
			{$x\text{-coordinate}=\liminf_{n\to\infty}\frac
			{\log\(-\log(\text{Block error probability})\)}
			{\log(\text{Block Length})}$};
		\draw(0,10/3.627)•node[left]{$\frac1\mu\approx.2757$}
			(5,0)•node[below]{$\beta=.5$};
		\draw[bad,dotted]
			(4.9,.001)•--+(0,.6)node[above,xshift=-40]
			{$\(.49,O(1)\)$ \cite[Theorem 1]{GX13}}
			(0,.01)•--+(.1,.1)node[above right,xshift=-10]
			{$\(O(\log n/n),O(1)\)$\cite[Theorem 1.6]{BGNRS18}};
		\draw[MHU16]plot[raw gnuplot]function{
			set parametric;
			mu=3.627;
			log2(x)=log(x)/log(2);
			h(x)=-x*log2(x)-(1-x)*log2(1-x);
			G(x)=1/(mu+1-mu*h(x));
			plot[0:.5]10*G(t)*t,10*(1-G(t))/mu smooth csplines;
			}(0,10/4.627)•node[left]{$\frac1{1+\mu}\approx.2161$}
			(1.2,1.3)node[below left]{Theorem \ref{thm:gamma}}
			(1.4,1.2)node[below left]{\cite[Theorem 7]{MHU16}};
		\draw[Our,dotted]
			(3.97561,.30494)•node[above right]{$(.3976,.0305)$}
				--(4.46955,0)•node[below]{$.4469$}
			(0,2.4)•--+(.6,.2)node[right]
				{$(0,.24)$ Section \ref{sec:obstacle}}
			(.1,2.4)•--+(1.1,-.2)node[right]
				{$(.01,.24)$ Section \ref{sec:2pocket}}
			(.2,2.4)•--+(.7,0)node[right]
				{$(.02,.24)$ Section \ref{sec:3pocket}}
			(0,1)•--+(.1,-.2)node[below right,xshift=-10]
				{$(0,.1)$ Section \ref{sec:obstacle}}
			(1.5,1)•--+(0,-.3)node[below]
				{$(.15,.1)$ Section \ref{sec:obstacle}};
		\draw[Our]plot coordinates{
			(0,10/3.627)
			(3.97560615940892, 0.304942511446948)
			(4.08687551008228, 0.241938247074105)
			(4.14395185864106, 0.212492019676073)
			(4.18848005771410, 0.190864774858869)
			(4.22639670594639, 0.173375279181456)
			(4.26004630944888, 0.158565973092125)
			(4.29064277330748, 0.145680286429480)
			(4.31891161042890, 0.134264910850402)
			(4.34532763048766, 0.124022438252995)
			(4.37022179406831, 0.114745044435068)
			(4.39383577845581, 0.106280589201336)
			(4.41635249705710, 0.0985136361705596)
			(4.43791438217050, 0.0913540709442582)
			(4.45863494254992, 0.0847298986167024)
			(4.47860639339929, 0.0785824839856242)
			(4.49790487683765, 0.0728632915246918)
			(4.51659414256492, 0.0675315845313478)
			(4.53472821008394, 0.0625527591887812)
			(4.55235333720022, 0.0578971114489503)
			(4.56950950380316, 0.0535389065774242)
			(4.58623154934287, 0.0494556651057623)
			(4.60255005798812, 0.0456276065853940)
			(4.61849205670940, 0.0420372104328875)
			(4.63408157247476, 0.0386688650283846)
			(4.64934008185885, 0.0355085842622566)
			(4.66428687740426, 0.0325437763126255)
			(4.67893936887402, 0.0297630533076069)
			(4.69331333299157, 0.0271560733600360)
			(4.70742312205696, 0.0247134084675097)
			(4.72128183942937, 0.0224264332692893)
			(4.73490148809463, 0.0202872307596161)
			(4.74829309720156, 0.0182885118921182)
			(4.76146683043664, 0.0164235466449761)
			(4.77443207932787, 0.0146861046044328)
			(4.78719754396506, 0.0130704035023392)
			(4.79977130315291, 0.0115710644389590)
			(4.81216087564151, 0.0101830727551690)
			(4.82437327378500, 0.00890174370310460)
			(4.83641505074225, 0.00772269221261710)
			(4.84829234215362, 0.00664180616477294)
			(4.86001090304469, 0.00565522269682278)
			(4.87157614063861, 0.00475930711003692)
			(4.88299314359456, 0.00395063405138396)
			(4.89426670814881, 0.00322597066850953)
			(4.90540136154829, 0.00258226149079616)
			(4.91640138311252, 0.00201661482394013)
			(4.92727082321111, 0.00152629047575281)
			(4.93801352040403, 0.00110868865647052)
			(4.94863311695878, 0.000761339917631317)
			(4.95913307292242, 0.000481896015981413)
			(4.96951667892789, 0.000268121589815320)
			(4.97978706784350, 0.000117886576083289)
			(4.98994722541618, 0.0000291592746097554)
		}(2.23,1.38)node[above right]{Theorem \ref{thm:main}};
	}

\clearpage

\subsection{Visualization of Polar Code Construction} \label{app:butterfly}

\newpage

	See also Section~\ref{sec:butterfly}.
	\fermat{Be careful}
	
	\tikz[rp,o,shift=(current page text area.south),yshift=\bbsize]{
		\let\E\expandafter
		\def\eatone#1{}
		\def\BIN#1#2{\pgfmathbin{32+#1}\edef#2{\E\eatone\pgfmathresult}}
		\foreach\j in{0,...,31}{\BIN\j\J\draw foreach\l in{1,...,6}{
				(BB cs:-\l,\j)node[E](\l E\J){}(BB cs:+\l,\j)node[D](\l D\J){}}
			(1E\J.NE)--(1D\J.NW)(1E\J.SE)--(1D\J.SW);}
		\def\readfive#1#2#3#4#5{\def\2{#2}\def\3{#3}\def\4{#4}\def\5{#5}}
		\def\BIN#1{\pgfmathbin{16+#1}\E\readfive\pgfmathresult}
		\def\DRAW#1#2{\foreach\j in{0,...,15}{\BIN\j\draw
			(#1E\k1.NW)--(#2E\k1.NE)(#1D\k1.NE)--(#2D\k1.NW)
			(#1E\k1.SW)--(#2E\k0.NE)(#1D\k1.SE)--(#2D\k0.NW)
			(#1E\k0.NW)--(#2E\k1.SE)(#1D\k0.NE)--(#2D\k1.SW)
			(#1E\k0.SW)--(#2E\k0.SE)(#1D\k0.SE)--(#2D\k0.SW);}}
		\def\k#1{\2\3\4\5#1}\DRAW12
		\def\k#1{\2\3\4#1\5}\DRAW23
		\def\k#1{\2\3#1\4\5}\DRAW34
		\def\k#1{\2#1\3\4\5}\DRAW45
		\def\k#1{#1\2\3\4\5}\DRAW56
	}

\end{document}